\documentclass[showpacs,preprintnumbers,amsmath]{revtex4}
\usepackage{epsfig}
\usepackage{graphicx}
\usepackage{bm}
\usepackage{amsmath}
\usepackage{amssymb}

\begin{document}

\title{Generalized Gibbs ensembles for time dependent processes}
\author{Ph. Chomaz}
 
\affiliation{GANIL, DSM-CEA/IN2P3-CNRS, 
BP 5027, F-14076\ CAEN cedex 5, FRANCE}

\author{F. Gulminelli} 
\altaffiliation{member of the Institut 
Universitaire de France}

\author{O.\ Juillet}

\affiliation{LPC Caen, IN2P3-CNRS et Universit\'{e},
F-14050 CAEN cedex, FRANCE}

\begin{abstract}
An information theory description of finite systems 
explicitly evolving in time is presented for classical 
as well as quantum mechanics. We impose a variational
principle on the Shannon entropy at a given time while 
the constraints are set at a former time. The resulting density 
matrix deviates from the Boltzmann kernel
and contains explicit time odd components which can be interpreted as 
collective flows. 
Applications include quantum brownian motion, linear response theory, 
out of equilibrium situations for which  
the relevant information is collected within different time 
scales before entropy saturation, and the dynamics of the 
expansion.  
\end{abstract}
\maketitle 
\newpage              

\section{\protect\smallskip Introduction}

The microscopic foundations of thermodynamics are well established using the
Gibbs hypothesis of statistical ensembles maximizing the Shannon
entropy\cite{jaynes}.When the thermodynamic limit can be taken, the various Gibbs
ensembles for infinite systems converge to a unique thermodynamic equilibrium
\footnote{%
For infinite systems the discussion is now focused mainly on the issue of
non-extensivity, linked to the introduction of
entropies different from the Shannon one\cite{tsallis}, 
and to the study of systems
with long range interactions\cite{dauxois}.}.
However, many systems studied in
physics do not correspond to this mathematical limit of infinite systems\cite
{hill} and, in fact, finite systems are now, per se, a subject of a very
intense research activity, from metallic clusters\cite{dauxois,clusters} to
Bose condensates\cite{bose,traps}, from nanoscopic systems\cite{quantum} to
atomic nuclei \cite{gross,high:energy} and elementary particles\cite{qgp}.
The question thus arises: can the equilibrium of a finite systems be defined. 

A priori, the Gibbs concept of statistical ensembles of replicas which is
applicable for an arbitrary number of particles, seems an ideal tool to
define the thermodynamics of finite systems. However, in a finite system the
various Gibbs ensembles are not equivalent\cite{inequiv} and
lead to different equilibria which
physical meaning and relevance has to be
investigated. In the different physical cases, which in the following
encompass both the case of isolated systems and of systems in contact with a
finite or infinite reservoir\footnote{%
When the system is finite the latter case deserves even more discussion since
the factorization of the thermodynamics of the bath is not guaranteed 
out of the
thermodynamic limit.}, the identification of the relevant statistical
ensemble is a key issue.

\smallskip From a macroscopic point of view, a common interpretation of a
statistical ensemble is an infinite collection of infinite subsystems of
the studied infinite system in a specific thermodynamic situation. The
required independence of the different subsystems is insured by the
thermodynamic limit. Within this interpretation, a single system can be
considered as a statistical ensemble and thus can be discussed in terms of
equilibrium. This kind of equilibrium is not relevant for a finite system
since i) the interface interactions between subsystem cannot be neglected,
ii) the procedure of coarse-graining modifies the entropic properties of the
system and iii) a finite system 
does not lead to an infinite ensemble of subsystems. In fact
a single realization of a finite system cannot be discussed in statistical physics 
terms.

An alternative viewpoint is given by the Boltzmann ergodic assumption. In
this interpretation the statistical ensemble represents the collection of
successive snapshots of a physical system evolving in time. The equivalence
between this time average and the Gibbs ensemble is then insured by the
ergodic theorem
\footnote{
For an isolated system described by an hamiltonian $H$, the ergodic picture
defines a unique microcanonical equilibrium characterized by all the
conserved quantities, the energy and other observable related to the
symetries of $H$ \cite{gross}. For a system in contact with a reservoir, we
may get the different statistical ensembles depending on the conserved
quantities characterizing the coupling with the bath.}. 
This interpretation however suffers from important drawbacks. First, not only a
proof of the ergodic hypothesis under fairly general conditions is lacking,
but even for a truly ergodic Hamiltonian, a finite time experiment may very
well achieve ergodicity only on a subspace of the total accessible phase
space\cite{thirring}. Moreover, ergodicity applies to confined systems and
thus it requires the definition of boundary conditions when the
thermodynamic limit does not apply. Then the statistical ensemble in general
explicitly
depends on the boundary conditions and we will discuss in this article
that an exact knowledge of the boundary corresponds to an infinite information
and is therefore hardly compatible 
with the very principles of statistical mechanics, as a reduction of the 
many body information to a (small) number of state variables. 

Finally, in many physical situation the ergodicity ideas do not apply.
Indeed, many physics experiments do not follow the time evolution of a
single system, but rather concern averages over a great number of events,
i.e. of physical replicas of systems experimentally prepared or sorted in
similar way which are then observed at a given time. In such a context
there is a priori no connection between the measuring time and the
time it takes for an ergodic system to visit evenly the energy shell.
Moreover, the systems experimentally accessible are often not confined but
freely evolve in the vacuum, as it is notably the case for atomic and
nuclear clusters and high energy heavy ion collisions. 
These open transient finite systems are isolated but
never stationary. The concept of a well defined stationary equilibrium,
uniquely defined by the variables conserved by the dynamics in an
hypothetical constraining box, is certainly not useful for these 
finite systems.

However statistical approaches, expressing the reduction of the available
information to a limited number of collective observables, can still be of
some pertinence for such complex many body systems \cite{balian}. The
physical meaning of such statistical ensembles is different from the case of
an ergodic system or a macroscopic coarse graining. The standard assumption
is that the dynamics is sufficiently complex (chaotic or mixing)\cite
{tolman,rasetti,cowley} such that, repeating an experiment many times, the
ensemble of events dynamically populates a phase space in a ''democratic''
enough way such that the gross features of the ensemble of events are
dominated by few collective variables. In this case, the maximum entropy
postulate cannot be justified from the ergodic theorem 
but has to be interpreted as a minimum information postulate which finds its
justification in the complexity of the dynamics independent of any time scale%
\cite{jaynes,balian}.

This information theory approach is a very powerful extension of the
classical Gibbs equilibrium: any arbitrary observable including time odd
quantities can act as a state variable, and all statistical quantities as
equations of state and phase diagrams can be unambiguously defined for any
number of particles\cite{noi}. The price to be paid for such a
generalization is that  the density matrix continuously
evolve in time as soon as the constraining observables 
are not restricted to conserved quantities, 
meaning that we have to take into account the time
dependence of the process. A well known example of application of the information theory to a dynamical process  is given by time dependent
mean field theories, which can be viewed as the solution of a maximum
entropy variational problem under the time dependent constraint of the
knowledge of all one body observables \cite{balian}.

In this article, we develop an information theory approach to the physics of
finite systems evolving in time. 
We show that such an approach provides a description of the 
thermalization process. 
In presence of non linear dynamics 
 it may lead to deviation from the Boltzmann-like 
exponential distributions and thus might provide an explanation for 
the appearance of non-extensive statistics\cite{tsallis}.  
In a mean-field 
context this dynamical extension envolves the RPA matrix 
leading to the introduction the collective degrees of freedom.     
For unconfined systems, we will show that
their finite size at a finite time can be accounted for by introducing, in
the Shannon information, additional constraints describing the system's
compactness (size and shape). This provides a natural solution to the
boundary condition problem. The same formalism describes 
incompletely known boundary conditions, 
confinement by a potential, and unconfined self bound or unbound systems.
%It unifies the formalism with the treatment of
%systems for which the confinement is insured by an external potential or by
%incompletely known boundary conditions. 
The time dependence of the process
naturally leads to the appearance of new time-odd constraints or collective
flows. In the case of an ideal gas of particles or clusters
%, or for interactions local in space as the Boltzmann collision integral,
%characterized by their mean square radius, 
we will show that the extension and flow constraints forms a closed 
algebra together with the  energy operator allowing an exact description of 
the free expansion of the system in vacuum.
%and that the information at the initial
%time is sufficient to calculate the exact density matrix at any successive
%time. 
%If the force is not zero ranged, the number of constraining 
%observables grows with time, and a complete description 
%can be obtained only by injecting the relevant information 
%at the time of entropy saturation.
 
\section{\protect\smallskip Statistical equilibria}

Let us first recall the standard formalism of statistical equilibria which we
will then extend to the case of time dependent processes involving finite
systems.

%\subsection{\protect\smallskip Gibbs formulation}

In order to describe a statistical ensemble, we %can 
introduce either the classical
density $\hat{D}\left( \vec{Q},\vec{P}\right) $ in the many-body phase
space associated with all the particle positions $\vec{Q}=\{q_i\}$ and
momenta $\vec{P}=\{p_i\}$ or the quantum density matrix $\hat{D}
=\sum_{\left( n\right) }\left| \Psi ^{\left( n\right) }\right\rangle
\;p^{\left( n\right) }\;\left\langle \Psi ^{\left( n\right) }\right| $ where 
$\left| \Psi ^{\left( n\right) }\right\rangle $ are the state of the
different events $\left( n\right) $ and $p^{\left( n\right) }$ the
associated occurrence probability. According to the Gibbs hypothesis,
equilibria are maxima of the Shannon entropy\cite{jaynes} 
\begin{equation}
S=-\mathrm{Tr}\hat{D}\log \hat{D},  \label{EQ:S}
\end{equation}
where $\mathrm{Tr}$ means an integral over the coarse-grained classical
phase space, $\left( \vec{Q},\vec{P}\right) $, or the trace over the
quantum Fock or Hilbert space of states $\left| \Psi \right\rangle $. In
this article we implicitly use units such that the Boltzmann constant $k=1$.

When the system is characterized by $L$ observables, $\vec{\hat{A}}=\{%
\hat{A}_{\ell }\}$, known in average $<\hat{A}_{\ell }>=\mathrm{Tr}\hat{D}%
\hat{A}_{\ell }$, one should maximize the constrained entropy

\[
S^{\prime }=S-\sum_{\ell }\lambda _{\ell }<\hat{A}_{\ell }> 
\]
where the $\vec{\lambda }=\{\lambda _{\ell }\}$ are $L$ Lagrange
multipliers associated with the $L$ constraints $<\hat{A}_{\ell }>$. The
Gibbs equilibrium
%, the maximum of the entropy under constraints, 
is then given by 
\begin{equation}
\hat{D}_{\vec{\lambda }}=\frac{1}{Z_{\vec{\lambda }}}\exp -\vec{
\lambda} .\vec{\hat{A}},  \label{EQ:D}
\end{equation}
where $\vec{\lambda} .\vec{\hat{A}}=\sum_{\ell =1}^{L}\lambda _{\ell }\hat{A}
_{\ell }$ and where $Z_{\vec{\lambda }}$ is the associated partition sum
insuring the normalization of $\hat{D}_{\vec{\lambda }}$.

To interpret the Gibbs ensemble as resulting from the contact with a
reservoir or to guarantee the stationarity of the equilibrium eq.(\ref{EQ:D}),
it is often assumed that the observables $\hat{A}_{\ell }$ are conserved
quantities such as the energy $\hat{H}$ or the particle (or charge) numbers
$\hat{N}_{i}$. However, 
%as we have already stressed in the introduction, ergodicity 
%is not the only path to equilibrium and
there is no formal reason to limit the state variables to constants of the
motion and, in fact, the introduction of non conserved quantities is a way
to take into account some non ergodic aspects. Indeed, an additional
constraint reduces the entropy, limiting the phase space and modifying the
event distribution. This point will be developed at length in the next
sections.

It should be noticed that the formalism recalled above encompass microcanonical thermodynamics\cite{gross} which 
can be obtained from the variation of the Shannon entropy 
eq.(\ref{EQ:S}) in a fixed energy subspace with no external constraints.
In this case the maximum of the Shannon entropy can be identified
with the Boltzmann entropy 
\[
max\left( S\right) =\log W\left( E\right) , 
\]
where $W$ is the total state density with the energy E. 

%Indeed 
%let us consider for a moment a sharp 
%energy constraint, $E^{(n)} - <\hat{H}>=0$ for all $(n)$. 
%the microcanonical
%density matrix 
%$\hat{D}_{E}$ maximizes the Shannon entropy and 
%corresponds to an even occupation of the whole energy shell
%\[
% S \left [\hat {D}_{E} \right ] = 
%\log W \left ( <\hat{H}>   \right ) = max,
%\] 
%while non ergodic components can be already included within the Gibbs 
%formalism through the introduction of extra constraints.  

The microcanonical
case can also be seen as a particular Gibbs equilibrium (\ref{EQ:D}) for
which both the energy and its fluctuation are constrained. This so called
Gaussian ensemble in fact interpolates between the canonical and
microcanonical ensemble depending upon the constraint on the energy
fluctuation\cite{gauss}. The same procedure can be applied to any
conservation law so that the Gibbs formulation (\ref{EQ:D}) can be
considered as the most general statistical ensemble.\

\section{Multiple time statistical ensembles}

\smallskip 

Eq. (\ref{EQ:D}) 
represents the standard statistical description of
a physical system at a given time. Indeed, if some of the
observables $\hat{A}_{\ell }$ are not constants of motion, then  
the statistical ensemble (\ref{EQ:D}) is not stationary, but
will evolve in time. 
This specific role played by time stresses
that this information theory ''equilibrium'' cannot be justified using the
usual ergodic arguments, 
but through the fact that the global features of the
replicas of the considered system are characterized, at a given time, by few
observables, i.e. that the information is concentrated in few degrees of
freedom.  
In many physical cases one can clearly identify a specific time 
("freeze out" time) 
at which the information concentrated in a given observable 
is frozen (i.e. the observable expectation 
value ceases to evolve or presents a trivial dynamics).
However this freeze out time may be fluctuating or  
different for different observables. For example for the ultra-relativistic
heavy ion reactions two freeze-out times are discussed, one for the chemistry
and one for the thermal agitation.  
To solve these questions we need to introduce time as an
explicit variable and define a statistical ensemble
constrained by informations coming from different times.
 
\subsection{\protect\smallskip Formulation of the issue}

Let us assume that the evolution of an ensemble can be written as 
\begin{equation}
\partial _{t}\hat{D}=\mathcal{F}[\hat{D}],  \label{EQ:Dyn}
\end{equation}
where $\mathcal{F}[\hat{D}]$ is a functional of the density matrix $\hat{D}.$
This is a very general dynamical evolution since it includes classical and
quantal Hamiltonian evolutions 
\begin{equation}
\partial _{t}\hat{D}=\{\hat{H},\hat{D}\},  \label{EQ:Liouville}
\end{equation}
where $\hat{H}$ is the system Hamiltonian and $\{.,.\}$ are Poisson bracket
in classical physics and commutators divided by $i\hbar $ in quantum
physics. Eq. (\ref{EQ:Dyn}) also includes non-linear approaches such as
mean-field approximations and more generally variational treatments for
which $\hat{H}$ is replaced by an effective operator which depends upon the
actual state $\hat{D}$ : $\hat{H}\rightarrow \hat{H}[\hat{D}].$ Eq. (\ref
{EQ:Dyn}) also includes the stochastic extensions of such approaches, $\hat{D}$ 
being the ensemble average of the stochastic evolutions.

Let us now suppose that the different informations on the system, $<\hat{A}
_{\ell }>$, are known at different times, $t_{\ell }$: 
\[
<\hat{A}_{\ell }>_{t_{\ell }}=\mathrm{Tr}\hat{D}\left( t_{\ell }\right) \hat{
A}_{\ell }. 
\]
A generalization of the Gibbs idea would be that at a time $t$ the least
biased state of the system corresponds to the maximum of the Shannon entropy,
considering all informations as constraints. Causality arguments imply that
this time $t$ should be larger or equal to all the $t_{\ell }.$ It should be
noticed that in the case of a Hamiltonian evolution (\ref{EQ:Liouville})
because the entropy is a constant of motion this remark has no implications
and the maximization can be performed at any time leading to the very same
result. %In this case the Maximum Entropy principle is indeed a constraint 
%on the initial conditions: only an initially equilibrated system can 
%be statistically treated in successive times.
In the case of dissipative systems the entropy grows and eventually saturates. 
%% at an asymptotic time when the
%% interactions can be neglected.
%are over (for short ranged interactions) or when the equations
%of motion can be exactly integrated (for long ranged ones). 
In this case the
Maximum Entropy principle has to be applied at the time of entropy
saturation (or the observation time if it occurs before).

The maximization of the entropy at time $t$ with the various constraints $<
\hat{A}_{\ell }>_{t_{\ell }}$ known at former times $t_{\ell }$ corresponds
to the free maximization of 
\begin{eqnarray}
S^{\prime } &=&S\left( t\right) -\sum_{\ell =1}^{L}\lambda _{\ell }<\hat{A}
_{\ell }>_{t_{\ell }}  \nonumber \\
S^{\prime } &=&-\mathrm{Tr}\left( \hat{D}\left( t\right) \log \hat{D}\left(
t\right) +\sum_{\ell =1}^{L}\lambda _{\ell }\hat{A}_{\ell }\hat{D}\left(
t_{\ell }\right) \right) ,  \label{EQ: S-Two-times}
\end{eqnarray}
where the $\lambda _{\ell }$ are the Lagrange parameters associated with all
the constraints. 
%This maximization will lead to a density matrix which can
%be considered as a generalized to time dependent processes of the Gibbs
%statistical ensembles (\ref{EQ:D}).

% *******************************************************************
% A discuter si on pense que c'est utile

% phrase re-ajoutée car elle continue a me sembler correcte - à discuter fran

%SupprimŽ car je ne pense pas que ce soit juste
%car la variation d'entropie vient de la propagation  donc 
%elle a aussi lieu si on a maximiisŽ au temps t0 phil

%% To understand the physical meaning of 
%% eq.(\ref{EQ: S-Two-times}) let us consider the 
%% simple case of a single observable $\hat{A}$ 
%% measured at time $t_{0}$. The density matrix $\hat{D}_{eq}$
%% corresponding to a standard Gibbs equilibrium maximizes the functional 
%% \[
%% S_{eq}^{\prime }=S(t_{0})-\lambda <\hat{A}>_{t_{0}}=
%% S^{\prime }-\Delta S
%% \]
%% where $\Delta S$ is the entropy increase from time $t_{0}$ to time $t$. 
%% The maximization of eq.(\ref{EQ: S-Two-times}) 
%% then leads to a density matrix at the measurement time $t_{0}$,
%% which is associated to an entropy
%% smaller than the equilibrium entropy, 
%% $S[D(t_{0})]\leq S[\hat{D}_{eq}]$.
%% The minimum biased density matrix defined by eq.(\ref{EQ: S-Two-times})
%% can be considered as the generalization of a Gibbs 
%% ensemble to out of equilibrium situations.

% *******************************************************************

\subsection{\protect\smallskip Minimum information under fluctuating-time
constraints}

Let us first assume that the various times only slightly differ by $\delta
t_{\ell }=t-t_{\ell }$, thus we can use  the equation of motion (\ref{EQ:Dyn}) to link 
the various times by
\[
\hat{D}\left( t_{\ell }\right) =\hat{D}\left( t\right) -\delta t_{\ell
}
\mathcal{F}[\hat{D}\left( t\right)]
%\partial _{t}\hat{D}\left( t\right) 
\]
in order to explicitly write the constrained entropy 
as a function of a unique density  $\hat{D}=\hat{D}\left( t\right) $
\smallskip 
\begin{equation}
S^{\prime }=-\mathrm{Tr}\left( \hat{D}\left( \log \hat{D}+\sum_{\ell
=1}^{L}\lambda _{\ell }\hat{A}_{\ell }\right) -\sum_{\ell =1}^{L}\delta
t_{\ell }\lambda _{\ell }\hat{A}_{\ell }\mathcal{F}[\hat{D}]\right) \ .
\label{EQ:Sprime}
\end{equation}
Computing the variation $\delta S^{\prime }$ of $S^{\prime }$ induced by a
modification of the density matrix $\hat{D}\rightarrow $ $\hat{D}+\delta 
\hat{D}$ we get
\[
\delta S^{\prime }=-\mathrm{Tr}\delta \hat{D}\left( \log \hat{D}%
+1+\sum_{\ell =1}^{L}\lambda _{\ell }\hat{A}_{\ell }+\sum_{\ell
=1}^{L}\lambda _{\ell }\delta t_{\ell }\hat{B}_{\ell }\mathcal{[}\hat{D}%
].\right) , 
\]
where the operator $\hat{B}_{\ell }$ are related to $\hat{A}_{\ell }$ and to
the functional derivative $\mathcal{M}_{[1,2]}[\hat{D}]=\partial \mathcal{F}_{\left[
1\right] }[\hat{D}]/\partial \hat{D}_{\left[ 2\right] }^{T}$ by 
\begin{equation}
\hat{B}_{\left[ 1\right] } [\hat{D}]
=-\mathrm{Tr}_{\left[ 2\right] }%
\hat{A}_{\left[ 2\right] }\mathcal{M}_{[2,1]}[\hat{D}],  \label{EQ:B}
\end{equation}
where the indices $\left[ i\right] $ indicate the space over which the
operators are acting and where $\hat{D}_{\left[ 1\right] }^{T}$ is the
transposed matrix when it applies, i.e. in quantum mechanics. In this case,
if we introduce a base $\left\{ \left| I\right\rangle \right\} $ to explicit
the trace, eq. (\ref{EQ:B}) reads: 
\begin{equation}
\hat{B}_{IJ}\mathcal{[}\hat{D}]=-\sum_{KL}\hat{A}_{KL}\frac{\partial 
\mathcal{F}_{LK}\mathcal{[}\hat{D}]}{\partial \hat{D}_{JI}}.
\end{equation}
The minimum biased density matrix (solution of $\delta S^{\prime }=0$) is
given by %the solution of the equation 
%\begin{equation}
%\log \hat{D}_{\vec{\lambda }}\left( t\right) =-1-\sum_{\ell
%=1}^{L}\lambda _{\ell }\hat{A}_{\ell }^{\prime }[\hat{D}_{\vec{\lambda }
%}\left( t\right) ],  \label{EQ:Equil}
%\end{equation}
\begin{equation}
\hat{D}_{\vec{\lambda }}\left( t\right) =\frac{1}{Z_{\vec{\lambda }%
}\left( t\right) }\exp -\sum_{\ell =1}^{L}\lambda _{\ell }\hat{A}_{\ell
}^{\prime }[\hat{D}_{\vec{\lambda }}\left( t\right) ],
\label{EQ:D-equil-fluct}
\end{equation}
where we have introduced a modified observable 
\begin{equation}
\hat{A}_{\ell }^{\prime }[\hat{D}_{\vec{\lambda }}\left( t\right)
]=\left( \hat{A}_{\ell }+
%%
%% \sum_{\ell =1}^{L}
%% Il restait une somme en trop !!!
%%
\delta t_{\ell }\hat{B}_{\ell }[%
\hat{D}_{\vec{\lambda }}\left( t\right) ]\right)  \label{EQ:A-prime}
\end{equation}
which takes into account the time difference between the various
observations. The associated Lagrange parameters are defined by the
equations of states 
\begin{equation}
<\hat{A}_{\ell }>_{t_{\ell }}=\mathrm{Tr}\left( \hat{D}_{\vec{\lambda }%
}\left( t\right) -\delta t_{\ell }\mathcal{F}[\hat{D}_{\vec{\lambda }%
}\left( t\right) ]\right) \hat{A}_{\ell },  \label{EQ:EOS}
\end{equation}
%This equations of states reduces $<\hat{A}_{\ell }>_{t_{\ell }}=<\hat{A}%
%_{\ell }^{\prime }>_{t}$only in the particular case of a linear dependance
%of $\mathcal{F}[\hat{D}]$ upon $\hat{D}.$ 
The partition sum is defined as 
\begin{equation}
Z_{\vec{\lambda }}\left( t\right) =\mathrm{Tr}\exp -\sum_{\ell
=1}^{L}\lambda _{\ell }\hat{A}_{\ell }^{\prime }[\hat{D}_{\vec{\lambda }%
}\left( t\right) ].  \label{EQ:Z}
\end{equation}
The equation $<\hat{A}_{\ell }^{\prime }>_{t}=-\partial LogZ_{
\vec{\lambda }}(t)/\partial \lambda _{\ell }$ always holds but a similar
relation between the equation of state (\ref{EQ:EOS}) and the partition
sum (\ref{EQ:Z}) is valid only if $\mathcal{F}[\hat{D}]$ is a linear functional
of $\hat{D}$. In such a case $<\hat{A}_{\ell }>_{t_{\ell }}=<\hat{A}_{\ell
}^{\prime }>_{t}$=$-\partial LogZ_{\vec{\lambda }}(t)/\partial \lambda
_{\ell }.$

The minimum information density matrix eq.(\ref{EQ:D-equil-fluct}) looks
like a standard Gibbs equilibrium but this formal analogy hides important
differences. Indeed the constraining observables are modified according to (%
\ref{EQ:A-prime}) which contains new operators $\hat{B}_{\ell }$. 
If the original observables $\hat{A}_{\ell }$ were time even, the operators 
$\hat{B}_{\ell }$ are time odd, and
therefore correspond to non stationary situations.

This can be more easily seen if we interpret eq. (\ref{EQ:D-equil-fluct}) as
an extended Gibbs equilibrium under the set of constraints 
$\{\hat{A}_{\ell},\hat{B}_{\ell }\}$%
\begin{equation}
\hat{D}_{\vec{\lambda}, \vec{\nu }}\left( t\right) =\frac{1}{Z_{\vec{\lambda}
,\vec{\nu }}\left( t\right) }\exp -\sum_{\ell =1}^{L}\lambda _{\ell }\hat{A}
_{\ell }-\sum_{\ell =1}^{L}\nu_{\ell}\hat{B}_{\ell}[\hat{D}_{\vec{\lambda},
\vec{\nu }}\left( t\right) ],
\end{equation}
where the parameters $\nu _{\ell }$ 
contains the time information
%% are given by 
$\nu _{\ell }=\delta
t_{\ell }\lambda _{\ell }$. 
%This demonstrates that the introduction of time
%fluctuations deeply transforms the statistical equilibrium.

Because of the possible $\hat{D}$ dependence of the $\hat{B}_{\ell }$
operators, the distribution $\hat{D}_{\vec{\lambda }}\left( t\right) $
might deviate significantly from the usual exponential behavior. This opens
an interesting possibility to encounter non Gibbsian statistics. In the
literature\cite{tsallis}, non Gibbsian information kernels are usually
derived from a modification of the entropy as, for example, in the case of a
Tsallis distribution. Here these anomalous statistics might be obtained from
the time dependence of the studied system. In the following sections we will 
elaborate more on this subject.

\subsection{\protect\smallskip Hamiltonian evolution}

Let us illustrate the above results in the case of a Hamiltonian evolution 
(\ref{EQ:Liouville}). Then the entropy is a constant of the motion $\dot{S}=-
\mathrm{Tr}\left( \{\hat{H},\hat{D}\left( t\right) \}(\log \hat{D}+1)\right)
=$ $\mathrm{Tr}\left( \hat{H}\{(\log \hat{D}+1),\hat{D}\left( t\right)
\}\right) =0$ so that the minimum biased trajectory is independent of the
time $t$ at which the entropy is maximized. Introducing 
\[
\hat{D}\left( t_{\ell }\right) =\hat{D}\left( t\right) -\delta t_{\ell }\{%
\hat{H},\hat{D}\left( t\right) \}, 
\]
and using the cyclic invariance of the trace (or the by part integration in
the phase space integral for the classical case), eq.(\ref{EQ:Sprime}) can
be written as

\begin{equation}
S^{\prime }=-\mathrm{Tr}\left( \hat{D}\left( \log \hat{D}+\sum_{\ell
=1}^{L}\lambda _{\ell }\hat{A}_{\ell }+\sum_{\ell =1}^{L}\delta t_{\ell
}\lambda _{\ell }\{\hat{H},\hat{A}_{\ell }\}\right) \right) \;.
\label{EQ:Sprime-reversible}
\end{equation}
In Eq. (\ref{EQ:Sprime-reversible}) we can introduce 
\[
\hat{A}_{\ell }^{\prime }=\hat{A}_{\ell }+\delta t_{\ell }\{\hat{H},\hat{A}
_{\ell }\} 
\]
in agreement with the Heisenberg picture, in which the observables would
evolve from time $t_{\ell }$ up to a common time $t$. The generalized
information theory result, i.e. the extremum of the variation $S^{\prime }$,
is given by 
\begin{equation}
\hat{D}_{\vec{\lambda }}\left( t\right) =\frac{1}{Z_{\vec{\lambda }
}\left( t\right) }\exp -\vec{\lambda} .\vec{\hat{A}}^{\prime }.
\label{EQ:Info-Equilibrium}
\end{equation}
In this case the $\hat{A}^{\prime }$ operators do not depend on $\hat{D}$ so
that the distribution (\ref{EQ:Info-Equilibrium}) is 
%% exponential 
similar to
a standard Gibbs equilibrium. The Lagrange parameter $\lambda _{\ell }$ are
defined by the constraints 
\[
<\hat{A}_{\ell }>_{t_{\ell }}=\mathrm{Tr}\hat{D}_{\vec{\ \lambda }}\left(
t_{\ell }\right) \hat{A}_{\ell }=\mathrm{Tr}\hat{D}_{\vec{\lambda }
}\left( t\right) \hat{A}_{\ell }^{\prime }=-\partial _{\lambda _{\ell }}\log
Z_{\vec{\lambda }}\left( t\right) . 
\]

Eq. (\ref{EQ:Info-Equilibrium}) can also be interpreted as the introduction
of additional constraints 
\[
\hat{B}_{\ell }=\{\hat{H},\hat{A}_{\ell }\} 
\]
and additional Lagrange parameter $\nu _{\ell }$ associated with the out of
equilibrium minimum biased density matrix 
\begin{equation}
\hat{D}_{\vec{\lambda ,\nu }}=\frac{1}{Z_{\vec{\lambda} ,\vec{\nu }}}\exp -%
\vec{\lambda} .\vec{\hat{A}}-\vec{\nu} .\vec{\hat{B}}.
\end{equation}
The equations of state are given by 
$<\hat{A}_{\ell }>_{t} =-\partial _{\lambda _{\ell }}\log Z_{\vec{\
\lambda} ,\vec{\nu }}\left( t\right) $ and 
$<\hat{B}_{\ell }>_{t} 
=-\partial _{\nu _{\ell }}\log Z_{\vec{\ \lambda}
,\vec{\nu }}\left( t\right)$ .
%
%\begin{eqnarray*}
%<\hat{A}_{\ell }>_{t} &=&\mathrm{Tr}\hat{D}_{\vec{\lambda} ,\vec{\nu }}\left(
%t\right) \hat{A}_{\ell }=-\partial _{\lambda _{\ell }}\log Z_{\vec{\
%\lambda} ,\vec{\nu }}\left( t\right) \\
%<\hat{B}_{\ell }>_{t} &=&\mathrm{Tr}\hat{D}_{\vec{\lambda} ,\vec{\nu }}\left(
%t\right) \hat{B}_{\ell }=-\partial _{\nu _{\ell }}\log Z_{\vec{\ \lambda}
%,\vec{\nu }}\left( t\right) .
%\end{eqnarray*}
Then the specific case (\ref{EQ:Info-Equilibrium}) is obtained requiring $%
\nu _{\ell }=\lambda _{\ell }\delta t_{\ell },$ the $\lambda $'s being
defined by the constraints 
$ %\[
<\hat{A}_{\ell }>_{t_{\ell }}=<\hat{A}_{\ell }>_{t}+\delta t_{\ell }<\hat{B}
_{\ell }>_{t}
%=-\partial _{\lambda _{\ell }}\log Z_{\vec{\lambda} ,\vec{\nu }
%}\left( t\right) -\delta t_{\ell }\log \partial _{\nu _{\ell }}Z_{\vec{\
%\lambda} ,\vec{\nu }}\left( t\right) . 
$. % \]

\subsection{\protect\smallskip Minimum information under multiple-time
constraints}

\label{multiple}

This procedure can be easily extended to longer time intervals. To do that,
we have to integrate the evolution from $t_{\ell }$ to $t$ 
\[
\hat{D}\left( t_{\ell }\right) =\hat{D}\left( t\right) -\int_{t_{\ell
}}^{t}dt^{\prime }\mathcal{F[}\hat{D}\left( t^{\prime }\right) ]. 
\]
The variations of the density matrix at various times are related by 
\[
\delta \hat{D}\left( t^{\prime }\right) =\delta \hat{D}\left( t\right)
-\int_{t^{\prime }}^{t}dt"\mathcal{F[}\hat{D}\left( t"\right) +\delta \hat{D}
\left( t"\right) ]-\mathcal{F[}\hat{D}\left( t"\right) ], 
\]
which leads to the relation
\begin{equation}
\delta \hat{D}_{[1]}\left( t^{\prime }\right) =\delta \hat{D}_{[1]}\left(
t\right) -\int_{t^{\prime }}^{t}dt^{"}\mathrm{Tr}_{[2]}\frac{\partial 
\mathcal{F}_{[1]}[\hat{D}\left( t"\right) ]}{\partial \hat{D}_{[2]}^{T}}
\delta \hat{D}_{[2]}\left( t"\right) ,  \label{EQ:delta-D-all}
\end{equation}
where the indices $\left[ i\right] $ indicates the space over which the
operators are acting (see equation (\ref{EQ:B}) ). This equation can be
integrated giving
\[
\delta \hat{D}_{[1]}\left( t^{\prime }\right) =\delta \hat{D}_{[1]}\left(
t\right) -\sum_{p=1}^{\infty }\frac{(t-t^{\prime })^{p}}{p!}\mathrm{Tr}_{[2]}%
\mathcal{M}_{[1,2]}^{(p)}\left( t,t^{\prime }\right) \delta \hat{D}%
_{[2]}\left( t\right) , 
\]
where the matrices $\mathcal{M}_{[1,2]}^{(p)}$ are defined iteratively by
\begin{eqnarray*}
%\mathcal{M}_{[1,2]}^{(1)}\left( t,t^{\prime }\right) &=&\frac{1}{%
%(t-t^{\prime })}\int_{t^{\prime }}^{t}dt^{"}\frac{\partial \mathcal{F}_{[1]}[%
%\hat{D}\left( t"\right) ]}{\partial \hat{D}_{[2]}^{T}}, \\
\mathcal{M}_{[1,2]}^{(p+1)}\left( t,t^{\prime }\right) &=&-\frac{p+1}{%
(t-t^{\prime })^{p+1}}\int_{t^{\prime }}^{t}dt^{"}\mathrm{Tr}_{[3]}\frac{%
\partial \mathcal{F}_{[1]}[\hat{D}\left( t"\right) ]}{\partial \hat{D}%
_{[3]}^{T}}\mathcal{M}_{[3,2]}^{(p)}\left( t,t^{"}\right) \left(
t-t^{"}\right) ^{p}.
\end{eqnarray*}
with $\mathcal{M}_{[1,2]}^{(0)}=1$. If we introduce this expression into the constrained entropy $S^{\prime }$
eq.(\ref{EQ: S-Two-times}) to compute the variation $\delta S^{\prime }$ 
%induced by a modification of the
%density matrix at the time $t$, $\hat{D}\left( t\right) \rightarrow $
% $\hat{D%}\left( t\right) +\delta \hat{D}\left( t\right) $ 
,we get
\[
\delta S^{\prime }=-\mathrm{Tr}\delta \hat{D}\left( t\right) \left( \log 
\hat{D}\left( t\right) +1+\sum_{\ell =1}^{L}\lambda _{\ell }\left( \hat{A}%
_{\ell }+\sum_{p=1}^{\infty }\frac{(t-t_{\ell })^{p}}{p!}\hat{B}_{\ell
}^{(p)}
\left( t,t_{\ell } \right) 
%[\hat{D}\left( t\right) ]
\right) \right) , 
\]
where the operator $\hat{B}_{\ell }^{(p)}$ are related to $\hat{A}_{\ell }$
and to the matrices $\mathcal{M}_{[1,2]}^{(p)}$ through 
\begin{equation}
\hat{B}_{\ell _{\left[ 1\right] }}^{(p)}
\left( t,t_{\ell } \right) 
%\mathcal{[}\hat{D}]
=-\mathrm{Tr}%
_{\left[ 2\right] }\hat{A}_{\ell _{\left[ 2\right] }}\mathcal{M}%
_{[2,1]}^{(p)}\left( t,t_{\ell }\right) .  \label{EQ:B-all}
\end{equation}
In the special case of a Hamiltonian evolution, the operators $\hat{B}
_{\ell }^{(p)}$ are simply $p-$uple commutators 
%\[
%\hat{B}^{(p)}=\{\hat{H},\hat{B}^{(p-1)}\}\;\;\;;\;\;\;\hat{B}^{(0)}=\hat{A}. 
%\]
$
\hat{B}^{(p)}=\{\hat{H},\hat{B}^{(p-1)}\}
$ with $\hat{B}^{(0)}=\hat{A}$. Therefore in this special case they do not depend
neither on time nor on $\hat{D}$. 
Coming back to the general case, 
as in eq.(\ref{EQ:A-prime}) above, we can introduce the new observable 
\[
\hat{A}_{\ell }^{\prime }
\left( t,t_{\ell } \right) 
%\mathcal{[}\hat{D}\left( t\right) ]
=\hat{A}_{\ell
}+\sum_{p=1}^{\infty }\frac{(t-t_{\ell })^{p}}{p!}\hat{B}_{\ell }^{(p)}
\left( t,t_{\ell } \right) 
%[\hat{D}\left( t\right) ]. 
\]
The minimum biased density matrix is
given by %the solution of the equation 
%\[
%\log \hat{D}_{\vec{\lambda }}\left( t\right) =-1-\sum_{\ell
%=1}^{L}\lambda _{\ell }\hat{A}_{\ell }^{\prime }[\hat{D}_{\vec{\lambda }
%}\left( t\right) ] 
%\]
%or
\begin{equation}
\hat{D}_{\vec{\lambda }}\left( t\right) =\frac{1}{Z_{\vec{\lambda }%
}\left( t\right) }\exp -\vec{\lambda}.\vec{\hat{A}}^{\prime }
\left( t,t_{\ell } \right) 
%[\hat{D}_{\vec{\lambda }}\left( t\right) ]
.  \label{EQ:D-equil-t-tprime}
\end{equation}
%The partition sum $Z_{\vec{\lambda }}\left( t\right) $ insures the
%normalization of $\hat{D}_{\vec{\lambda }}\left( t\right) $ 
%\[
%Z_{\vec{\lambda }}\left( t\right) =\mathrm{Tr}\exp -\vec{\ \lambda .
%\hat{A}}^{\prime }[\hat{D}_{\vec{\lambda }}\left( t\right) ] 
%\]
where the partition sum 
$Z_{\vec{\lambda }}\left( t\right)
=\mathrm{Tr}\exp -\vec{\lambda} .
\vec{\hat{A}}^{\prime }
\left( t,t_{\ell } \right) 
%[\hat{D}_{\vec{\lambda }}\left( t\right) ] 
 $ 
 insures the
normalization of $\hat{D}_{\vec{\lambda }}\left( t\right) $. 
The Lagrange parameter $\lambda _{\ell }$ are defined by the constraints 
\begin{equation}
<\hat{A}_{\ell }>_{t_{\ell }}=\mathrm{Tr}%
\left( \hat{D}_{\vec{\lambda }}\left( t\right) -\int_{t_{\ell
}}^{t}dt^{\prime }\mathcal{\ F[}\hat{D}_{\vec{\lambda }}\left( t^{\prime
}\right) ]\right) \hat{A}_{\ell },  \label{eos_general}
\end{equation}
which only for a linear dependence of $\partial _{t}\hat{D}$ on $\hat{D}$
such as in the particular case of a Hamiltonian evolution give back the
standard expression for the equations of state 
$
<\hat{A}_{\ell }>_{t_{\ell }}=
-\partial _{\lambda _{\ell }}\log Z_{
\vec{\lambda }}\left( t\right) 
$ since in this case 
$
<\hat{A}_{\ell }>_{t_{\ell }}%
%=\mathrm{Tr}\hat{D}_{\vec{\lambda }}\left( t_{\ell }\right) \hat{A}_{\ell }
=\mathrm{Tr}\hat{D}_{\vec{\lambda }}\left( t\right) \hat{A}_{\ell
}^{\prime }\left( t,t_{\ell }\right)
=<\hat{A}_{\ell }^{\prime }>_{t}
$.
%
%\[
%<\hat{A}_{\ell }>_{t_{\ell }}%
%%=\mathrm{Tr}\hat{D}_{\vec{\lambda }}\left( t_{\ell }\right) \hat{A}_{\ell }
%=\mathrm{Tr}\hat{D}_{\vec{\lambda }}\left( t\right) \hat{A}_{\ell
%}^{\prime }\left( t-t_{\ell }\right) =-\partial _{\lambda _{\ell }}\log Z_{%
%\vec{\lambda }}\left( t\right) . 
%\]
%% It is easy to verify that in the Hamiltonian case the time evolution of
%% the averages is given by 
%% \[
%% \partial _{t}<\hat{A}>=-<\{\hat{H},\hat{A}\}>. 
%% \]
It is interesting to note that if 
%% (and only if) 
the evolution is
Hamiltonian, the $\hat{A}_{\ell }^{\prime }$ represent the time evolution
of the constraining observables $\hat{A}_{\ell }$ in the Heisenberg
representation 
$\hat{A}_{\ell }^{\prime }
\left( t,t_{\ell }\right)
%\mathcal{[}\hat{D}\left( t\right) ]
=
\hat{A}_{\ell }^{\prime }\left( \Delta t_{\ell }\right) =e^{-i\Delta t_{\ell
}\hat{H}}\hat{A}_{\ell }e^{i\Delta t_{\ell }\hat{H}}$
%=\sum_{p=0}^{\infty }\frac{\Delta t^{p}}{p!}\hat{B}
%^{(p)} 
where $\Delta t_{\ell }=t-t_{\ell }.$ 

Coming back to the general case, 
as for eq.(\ref{EQ:D-equil-fluct}) above, eq. (\ref{EQ:Info-Equilibrium})
can also be interpreted as the introduction of additional constraints 
$\hat{B}_{\ell }^{(p)}$ and additional Lagrange parameters 
$\nu _{\ell }^{(p)}$
associated with the time evolution of the system 
\begin{equation}
\hat{D}_{\vec{\lambda} ,\vec{\nu }}
=\frac{1}{Z_{\vec{\lambda} ,\vec{\nu }}}\exp -%
\vec{\lambda} .\hat{A}-\sum_{p=1}^{\infty }\vec{\nu }^{(p)}.\vec{
\hat{B}}^{(p)}
%[\hat{D}_{\vec{\lambda} ,\vec{\nu }}]
,  \label{multistep}
\end{equation}
%where the partition sum $Z_{\vec{\lambda ,\{\nu \}}}\left( t\right) $
%insures the normalization of $\hat{D}_{\vec{\lambda ,\{\nu \}}}\left(
%t\right) $ 
%\[
%Z_{\vec{\lambda ,\{\nu \}}}\left( t\right) =\mathrm{Tr}\exp -\vec{\
%\lambda .\hat{A}-}\sum_{p=1}^{\infty }\vec{\nu }^{(p)}\vec{.\hat{B}}
%^{(p)}[\hat{D}_{\vec{\lambda ,\{\nu \}}}] 
%\]
%and 
where the Lagrange parameter $\nu _{\ell }^{(p)}$ are related to $\lambda
_{\ell }$ by $\nu _{\ell }^{(p)}=\frac{(t-t_{\ell })^{p}}{p!}\lambda _{\ell
} $ .

It is important to notice that 
Eqs. (\ref{EQ:D-equil-t-tprime}) or (\ref{multistep}) provide exact solution of
the complete many body evolution problem eq.(\ref{EQ:Dyn}) with a minimum
information hypothesis on the final time $t$ having made a set of 
observations $<\hat{A}_{\ell }>$ at previous times $t_{\ell }$.
%This remark shows the wide domain
%of applicability of information theory. 
We can see from eq.(\ref{multistep})
that in general an infinite amount of information, i.e. an infinite number
of Lagrange multipliers are needed if we want to follow the evolution of the
density matrix for a long time, or if the time interval between each
constraint is long.
% after the knowledge of the few observables $<%
%\hat{A}_{\ell }>$. 
%This clearly shows the complexity of the many body problem, 
%and nicely expresses the fact that a general and exact solution 
%of the many body evolution dynamics is essentially out of reach 
%even in the classical world.
%
However, 
%we can also see from eq.(\ref{multistep}) that 
if we are only
interested in short time scales the series will rapidly converge. Moreover
in the next sections we will show that different interesting physical
situations exist, for which the series can be analytically summed up. In
this case, a limited information (the knowledge of a small number of average
observables) will be sufficient to describe the whole density matrix at any
time, under the unique hypothesis that the information was finite at a given
time.

In the next sections we will illustrate this theory
with two representative examples: a dissipative non-Hamiltonian dynamics 
and a self consistent case. 

\subsection{\protect\smallskip Application: Brownian motion}
\label{brown} 

%Let us illustrate the theory formulated in the previous
%section with two important examples a %applications.
%The Boltzmann collision integral is examined in 
%section \ref{coll} while an example 
%dissipative non-Hamiltonian dynamics and a self consistent case. 
%is given in section \ref{brown}.
%In both cases we will show that the energy dissipation can be 
%accounted for at any time by introducing a finite number of extra
%constraints in a generalized Gibbs density matrix.
%\subsection{\protect\smallskip Boltzmann collision integral}\label{coll}
%\subsection{\protect\smallskip Brownian motion}\label{brown}

The description of the dynamics of quantum systems interacting with their
environment is, in general, a very difficult task due to the
system-reservoir interaction that usually involves a huge or even an
infinite number of degrees of freedom. The Liouville-von Neumann equation
for the total closed system is therefore useless for a reasonable
description. The standard prescription is then to construct effective
equations of motion for the reduced density matrix of the system by tracing
over the environmental variables in the exact dynamics. The resulting
quantum master equation includes dissipative and stochastic terms that take
into account the irrelevant degrees of freedom in an approximate way.

One prototype of system-reservoir models is the Caldeira-Leggett model \cite
{caldeira} for a Brownian particle of mass $m$, with coordinate ${\hat{r}}$
and momentum ${\hat{p}}$ , in a bath consisting of a large number of
harmonic oscillators. After performing a series of approximations\cite
{caldeira}, the density matrix of the Brownian particle is found to satisfy
the following Liouvillian quantum master equation :

\begin{equation}
\partial _{t}\hat{D}=\mathcal{F}[\hat{D}]=\frac{1}{i\hbar }\left[ \frac{{%
\hat{p}}^{2}}{2m},\hat{D}\right] -\frac{i\gamma }{\hbar }\left[ {\hat{r}}
,\left[ {\hat{p}},\hat{D}\right] _{+}\right] -\frac{d}{\hbar ^{2}}\left[ {%
\hat{r}},\left[ {\hat{r}},\hat{D}\right] \right]  \label{EQ:brown}
\end{equation}
The first term describes the free Hamiltonian dynamics. The second term,
proportional to the friction coefficient $\gamma $, is a dissipative term
associated to the mean coupling to the bath. The last term, with the
diffusion coefficient $d$ satisfying the Einstein relation $d=2m\gamma
/\beta $ where $\beta ^{-1}$ is the bath temperature, describes thermal
fluctuations.

Let us apply the general formalism of section \ref{multiple} to a one
dimensional Brownian particle prepared, at a time $t_{0}$ , with a mean
kinetic energy $<\hat{K}>_{t_{0}}=<\frac{{\hat{p}}^{2}}{2m}>_{t_{0}}$, and
which is observed at a later time $t$. Following the procedure of section 
\ref{multiple}, the information theory ansatz for the density matrix $\hat{D}
(t)$ is a generalized statistical equilibrium with the constraining
observables $\hat{B}^{(0)}=\hat{K}$ and $\hat{B}_{[1]}^{(p)}=-\mathrm{Tr}
_{\left[ 2\right] }\hat{B}_{\left[ 2\right] }^{(p-1)}\partial \mathcal{F}
_{\left[ 2\right] }/\partial \hat{D}_{\left[ 1\right] }^{T}$. With the
irreversible dynamics (\ref{EQ:brown}), we easily show that
\begin{eqnarray}
\hat{B}^{(1)} &=&4\gamma \hat{K}-\frac{d}{m} \\
\hat{B}^{(p)} &=&4\gamma \hat{B}^{(p-1)}=\left( 4\gamma \right) ^{p-1}\hat{B}%
^{(1)}
\end{eqnarray}
Using (\ref{EQ:D-equil-t-tprime}), the density matrix $\hat{D}(t)$ can be
written as 
\[
\hat{D}_{\lambda }\left( t\right) =\frac{1}{Z_{\lambda }^{\prime }\left(
t\right) }\exp -\lambda \hat{K}^{\prime }\left( t\right) , 
\]
where $Z_{\lambda }^{\prime }\left( t\right) 
=\mathrm{Tr}\exp -\lambda \hat{K}^{\prime }\left( t\right) \ $
and where the modified constraining operator 
$\hat{K}^{\prime }\left( t\right) $ can be actually resumed as 
\begin{equation}
\hat{K}^{\prime }(t)
%\mathcal{[}\hat{D}\left( t\right) ]
=\sum_{p=0}^{\infty }%
\frac{(t-t_{0})^{p}}{p!}\hat{B}^{(p)}
%[\hat{D}\left( t\right) ]
=e^{4\gamma
\left( t-t_{0}\right) }\hat{K}-\frac{d}{4m\gamma }\left( e^{4\gamma \left(
t-t_{0}\right) }-1\right) .  \label{EQ:Kprime}
\end{equation}
The Lagrange parameter $\lambda $ fulfills  the equation of state 
$<\hat{K}^{\prime }>_{t}=-\partial _{\lambda }\log Z_{\lambda }^{\prime
}\left( t\right) $ which gives after an explicit calculation of the
partition sum 
\begin{equation}
\frac{1}{2\lambda }=e^{4\gamma \left( t-t_{0}\right) }<\hat{K}>_{t}. 
\label{lambda_brown}
\end{equation}
%
%Therefore, $\lambda $ is implicitly time dependent. 
Since the evolution
Kernel $\mathcal{F}[\hat{D}]$ is a linear functional of $\hat{D},$ the
relation $<\hat{K}^{\prime }>_{t}=<\hat{K}>_{t_{0}}$holds 
(see eq.(\ref{eos_general})). 
Taking advantage of the explicit
expression (\ref{EQ:Kprime}) of the observable $\hat{K}^{\prime }(t)$ we
get 
\begin{equation}
%\frac{1}{2\lambda e^{4\gamma \left( t-t_{0}\right) }}=
<\hat{K}>_{t}=<\hat{K}%
>_{t_{0}}e^{-4\gamma \left( t-t_{0}\right) }+\frac{d}{4m\gamma }\left(
1-e^{-4\gamma \left( t-t_{0}\right) }\right) 
\label{eos_brown}
\end{equation}
This expression corresponds to the exact mean
value of the kinetic term at time $t$, as deduced from the equation of
motion of $<\hat{K}>$ induced by the quantum master equation (\ref{EQ:brown}
) 
\[
\frac{d<\hat{K}>}{dt}=-4\gamma <\hat{K}>+\frac{d}{m} 
\]
which clearly shows the damping role played by $\gamma $ and the fluctuation
ensure by $d$. 

Let us now interpret the above results. Since $\hat{K}^{\prime }\left(
t\right) $ is a combination of the unit and kinetic operator, the density
matrix can be recasted as 
\[
\hat{D}_{\lambda (t)}\left( t\right) =\frac{1}{Z_{\lambda (t)\nu \left(
t\right) }}\exp -\lambda (t)\nu \left( t\right) \hat{K}, 
\]
with $\nu (t)=exp\left( 4\gamma \left( t-t_{0}\right) \right) $. 
In the above equation we have make explicit the fact that  
the Lagrange multiplier $\lambda$ is time dependent as can be seen 
solving Eq. (\ref{lambda_brown})  with the help of Eq. \ref{eos_brown}.  
 This
density matrix can be interpreted as a canonical equilibrium with a time
dependent temperature defined by
$T^{-1}\left( t\right)=\lambda (t)\nu\left( t\right)$. 
Indeed  the equation of state defining $\lambda (t)$ 
implies $T\left( t\right) =2<\hat{K}>_{t}$ which is 
the standard result for an ideal 1D gas. 
Using eq.(\ref{eos_brown}) we can work out the time dependence of the
temperature
\[
T\left( t\right) =2<\hat{K}>_{t_{0}}e^{-4\gamma \left(
t-t_{0}\right) }+\beta ^{-1}\left( 1-e^{-4\gamma \left( t-t_{0}\right)
}\right) 
\]
$T(t)$ exponentially relaxes from the initial
temperature $2<\hat{K}>_{t_{0}}$to the bath temperature $\beta ^{-1}$ with
the correct characteristic time $1/4\gamma $, in agreement with 
the exact dynamics\cite{caldeira}.

Thus applying the generalized Gibbs ensembles for time dependent processes
to the problem of a quantum particle in contact with a reservoir we have
been able to describe the general evolution as a succession of canonical
statistical ensembles correctly relaxing toward the expected equilibrium.
This exemple illustrates the potential of the developed formalism to
statistically describe time dependent out-of-equilibrium phenomena.

\smallskip

\subsection{\protect\smallskip Application: Mean-field and non-linear dynamics}
\label{rpa} 

As an illustration of a non-linear dynamics, let us study the case of
self-consistent approaches which are of the mean-field type. To be more
specific, we consider a quantum system characterized by its one body density $%
\hat{\rho}$ , which can either be seen as the density matrix projected over all
particles but one ($\hat{\rho}_{[1]}=\mathrm{Tr}_{[2,...,N]}\,\hat{D}
_{[1,2,...,N]}$) or the expectation value of a generic one body operator 
$\hat{\rho}_{I,J}=<a_{J}^{+}a_{I}>$ where $a_{J}^{+}(a_{I})$ are
creation(anihilation) operators of a particle in the orbital $J(I)$. Since
we are studying a one-body approach we will restrict the discussion to
one-body operators $\hat{A}_{\ell }.$  

The time dependent mean-field dynamics is given by 
\begin{equation}
\partial _{t}\hat{\rho}=\{\hat{W}[\hat{\rho}],\hat{\rho}\},
\label{EQ:mean-field}
\end{equation}
where $\hat{W}[\hat{\rho}]$ is the self-consistent mean-field Hamiltonian.
Using a variational approach\cite{balian} 
$\hat{W}[\hat{\rho}]$ can be related to the
functional derivative of the energy $E=<\hat{H}>$ : $\hat{W}[\hat{\rho}
]_{[1]}=\partial E/\partial \hat{\rho}_{\left[ 1\right] }^{T}$ i.e. $\hat{W}[
\hat{\rho}]_{IJ}=\partial E/\partial \hat{\rho}_{JI}^{T}$. 
If we consider a small deviation from an equilibrium solution,
$\hat{\rho}=\hat{\rho}_0+\delta \hat{\rho}$
the dynamics of $\delta \hat{\rho}$ follows the time
dependent linear response (RPA) equation\cite{ring} 
\begin{equation}
\partial _{t}\delta \hat{\rho}_{[1]}=\{\hat{W}[\hat{\rho}],\delta \hat{\rho}%
\}_{[1]}+\left\{ \sum_{\left[ 2\right] }\hat{V}_{\left[ 1,2\right] }[\hat{%
\rho}]\delta \hat{\rho}_{[2]},\hat{\rho}\right\} _{[1]},  \label{EQ:RPA}
\end{equation}
where $\hat{V}_{\left[ 1,2\right] }=\partial \hat{W}[\hat{\rho}]_{\left[
1\right] }/\partial \hat{\rho}_{\left[ 2\right] }^{T},$ i.e. $\hat{V}
_{IL,JK}[\hat{\rho}]=\partial \hat{W}[\hat{\rho}]_{IJ}/\partial \hat{\rho}
_{KL},$ is interpreted as the residual interaction. 
It is more convenient to introduce the Liouville space, considering the density
matrices $\hat{\rho}$ as vectors $\parallel \rho \gg $ of components $\alpha
=(i,j)$ and one body operators $\hat{A}$ as dual vectors $\ll A\parallel $
using the scalar product $\ll A\parallel \rho \gg =\mathrm{Tr}\hat{A}^{+}
\hat{\rho}$ 
\footnote{It should be noticed that in the Liouville 
space the measure is simply a projection $<\hat{A}>=$ 
$\ll A\parallel \rho \gg .$}.
In this representation the dynamics of $\delta \hat{\rho}$ can be
written\cite{frascaria} 
\begin{equation}
i\partial _{t}\parallel \delta \rho \gg =\mathcal{H}\parallel \delta \rho
\gg =\left( \mathcal{W}-\mathcal{N\,V}\right) \parallel \delta \rho \gg
\label{rpa:2}
\end{equation}
where the RPA matrix $\mathcal{H}$ has a commutator structure
%\footnote{
%To show the hermiticity one should go back to the definition $\ll A\parallel 
%\mathcal{B}^{+}\parallel C\gg =\left( \ll C\parallel \mathcal{B}\parallel
%A\gg \right) ^{*}.$ In the special case of $\mathcal{W}$ and $\mathcal{N}$ ,
%using $\left( TrC^{+}\left[ B,A\right] \right) ^{*}=TrA\left[ B,C\right] $
%valide when $B$ is hermitian $B^{+}=B$ we get $\ll A\parallel \mathcal{B}
%^{+}\parallel C\gg =\ll A\parallel \mathcal{B}\parallel C\gg .$ As far as
%the residual interaction using the hermiticity of $\hat{W}$ and its
%definition we can easyly get an explicitly hermitian form $\mathcal{V}
%_{[1,2]}=\partial ^{2}E/\partial \hat{\rho}_{\left[ 1\right] }^{*}\partial 
%\hat{\rho}_{\left[ 2\right] }^{T}.$} 
$\mathcal{W}\parallel \cdot\gg =\parallel \left[ W,\cdot \right] \gg ,$ 
$\mathcal{N}\parallel \cdot \gg=$ $\parallel \left[ \rho ,\cdot \right] \gg $ 
and $\mathcal{V}\parallel \cdot \gg =\parallel \frac{\partial W}
{\partial \rho ^{T}}\cdot \gg .$

It is easy to verify that the matrix 
$\mathcal{M}_{[1,2]}=\partial \mathcal{F}_{\left[ 1\right] }
/\partial \hat{\rho}_{\left[2\right] }^{T}$ 
introduced in section \ref{multiple} is $-i$-times the RPA matrix $\mathcal{H}$
\[
i\mathcal{M}_{IJ,KL}=\mathcal{H}_{IJ,KL}
=\hat{W}_{IK}\delta _{LJ}-\delta _{IL}\hat{W}_{KJ}+\hat{V}%
_{IL,MK}\hat{\rho}_{MJ}-\hat{\rho}_{IM}\hat{V}_{ML,JK} 
\]
while the new operators $\hat{B}_{\ell }$ introduced
to take into account the propagation of
the information $<\hat{A}_{\ell }>$ over the time interval $\delta t_{\ell
}=t-t_{\ell }$ are given by
\begin{equation}
\ll B_{\ell }\parallel =i\ll A_{\ell }\parallel \mathcal{H=}i\ll A_{\ell}
\parallel \mathcal{(W-NV)}
\end{equation}
To get a deeper insight into this relation, let us introduce the energies 
$\omega _{\nu }$ and eigenstates $\delta \hat{\rho}_{\nu }$ of the RPA matrix
\footnote{Recall that if the unperturbed solution corresponds to a minimum 
in the energy surface and not to a maximum or to a saddle point, then the 
RPA eigenmodes are real\cite{ring}.} 
\begin{eqnarray}
\omega _{\nu }\parallel \delta \rho _{\nu }\gg &=&\mathcal{H}\parallel \delta
\rho _{\nu }\gg \\
\ll C_{\upsilon }\parallel \mathcal{H} &=&\omega_{\nu }\ll 
C_{\upsilon }\parallel 
\end{eqnarray}
where $\ll C_{\upsilon}\parallel $ represents the dual basis
$\ll C_{\upsilon ^{\prime }}\parallel \delta \rho
_{\nu }\gg =\delta _{\nu ^{\prime }\nu }$. 
%and the associated closure
%relation $\sum_{\nu }\parallel \delta \rho _{\nu }\gg \ll C_{\upsilon
%}\parallel =1.$ The relation $\ll C_{\upsilon }\parallel \mathcal{H=}\omega
%_{\nu }\ll C_{\upsilon }\parallel $ is thus easyly demonstrated using the
%above relations. Then $\omega _{\nu }^{*}\parallel C_{\nu }\gg =$ $\mathcal{H%
%}^{+}\parallel C_{\nu }\gg $ also holds showing that, when $\left[ W,\rho
%\right] =0,$ $\hat{C}_{\nu }=\hat{Q}_{\upsilon }$ if $\omega _{\nu }$ is
%real.
%
The operators $\hat{B}_{\ell }$ can be expressed as
\begin{equation}
\ll B_{\ell }\parallel =i\sum_{\nu }\omega _{\nu }\ll A_{\ell }\parallel
\delta \hat{\rho}_{\nu }\gg \ll C_{\upsilon }\parallel
\end{equation}
Expanding the $\ll A_{\ell }\parallel $ over the same basis
\begin{equation}
\ll A_{\ell }\parallel =\sum_{\nu }\ll A_{\ell }\parallel \delta \hat{\rho}%
_{\nu }\gg \ll C_{\upsilon }\parallel
\end{equation}
we observe that $\hat{A}_{\ell }$ and $\hat{B}_{\ell }$ are akin to
conjugated operators. To make this relation more explicit let us assume
that the operator $\hat{A}_{\ell }$ excites mainly one state $\nu _{\ell }$
with a real frequency $\omega _{\nu }$ with an excitation amplitude $\ll
A_{\ell }\parallel \delta \hat{\rho}_{\nu _{\ell }}\gg =a_{\nu _{\ell }}$.
Taking care of the fact that RPA eigenstates appears by pairs\cite{ring}, 
$\hat{A}_{\ell }$ and $\hat{B}_{\ell }$ can be written as (see appendix C for
details)
\begin{eqnarray}
\hat{A}_{\ell }&=&a_{\upsilon _{\ell }}\hat{C}_{\upsilon _{\ell }}+a_{\upsilon
_{\ell }}^{*}\hat{C}_{\upsilon _{\ell }}^{+} \\
 \hat{B}_{\ell }&=&i\omega _{\nu _{\ell }}a_{\upsilon _{\ell }}\hat{C}%
_{\upsilon _{\ell }}-i\omega _{\nu _{\ell }}a_{\upsilon _{\ell }}^{*}\hat{C}%
_{\upsilon _{\ell }}^{+}
\end{eqnarray}
If $\hat{A}_{\ell }$ is interpreted as a collective coordinate associated
with the creation of a collective mode through the $\hat{C}_{\upsilon _{\ell
}}^{+}$ operator, then the extra constraint
$\hat{B}_{\ell }$ taking care of the time dependence has the
structure of the associated momentum. 

It is interesting to notice that if the RPA
matrix does not evolve in time, the above treatment 
can be iterated to propagations over a finite time $t$.   
In this case $\ll B_{\ell}^{(p+1)}\parallel 
=i\ll B_{\ell }^{(p)}\parallel \mathcal{H}$ so that 
if we consider a single mode  
\begin{eqnarray}
\hat{B}_{\ell }^{(2p)} &=&\left( -\omega _{\nu _{\ell }}^{2}\right) ^{p}\hat{%
A}_{\ell } \\
\hat{B}_{\ell }^{(2p+1)} &=&\left( -\omega _{\nu _{\ell }}^{2}\right) ^{p}%
\hat{B}_{\ell }
\end{eqnarray}
then the time dependence of the process can be accounted through
a simple time dependence of the constraining operator 
\[
\hat{A}_{\ell }^{\prime }\left( t\right) =\hat{A}_{\ell }\cos \omega _{\nu
_{\ell }}(t-t_{\ell })+\hat{B}_{\ell }\omega _{\nu _{\ell }}^{-1}\sin \omega
_{\nu _{\ell }}(t-t_{\ell }).
\]
which represents an oscillation in the RPA collective phase space 
$\hat{A}_{\ell }$ and $\hat{B}_{\ell }.$ 

Thus applying the proposed formalism 
to non-linear dynamics we have
shown that the Gibbs ensembles description for time dependent processes
can take into account new degrees of freedom. 
In the mean-field example,   the 
RPA modes appear naturally and the 
generalized ensemble correctly describe 
their collective oscillation. 
This demonstrates how the developed formalism 
allows to go beyond the standard approaches 
to take into account dynamical aspects in a statistical 
manner.

\section{\protect\smallskip Unbound systems and the dynamics of the expansion}
\label{unbound}

An interesting application of the developed formalism is given by the case of finite
systems in the presence of a continuum. For such physical systems
the boundary condition problem is far from being trivial as we discuss 
below. We will show that the only consistent way to describe the boundary 
of unbound systems is to introduce in the Gibbs equilibrium 
at least one observable related to the system's extension (size and shape). 
Since these observables are not constants of the motion, 
such states are not stationary: in the absence of a confining force, 
the size of the system increases with time and collective flows are generated.
An especially relevant physical case is given by 
%high energy heavy ion 
nuclear collisions
\cite{high:energy,low:energy}  which produce a transient
correlated high density state that subsequently freely expands in the vacuum.
We will show that the out of equilibrium expansion can be 
%exactly 
accounted
for at any time within a finite amount of information, with the natural
emergence of time odd collective flow observables among the state variables.

\subsection{\protect\smallskip Boundary condition problem}
\label{boundary}

The information theory formalism is valid for any system
size and thus can be, a priori, applied to finite systems. However, as soon
as one $\hat{A}_{\ell }$ (e.g. $\hat{H}$) contains differential operators such as a kinetic
energy, the standard Gibbs equilibrium eq. (\ref{EQ:D}) is is a priori not defined,
unless boundary conditions are specified. 
The same is true for microcanonical thermodynamics, since
 this latter 
can be seen as a special case of the more general Gibbs formalism.
Boundary conditions are irrelevant at the thermodynamic limit (when
it exists), or if the system is 
%self 
bound: in both cases the definition
of a boundary corresponds to negligeable surface effects. 
However, in a finite system in the presence of a continuum this is never the
case, and in this respect the statistical physics of small unbound systems is
ill-defined.

\subsubsection{Systems in an external potential}
\label{harmonic}

Physical situations occur in which an external potential is used to confine
the system, as for recent studies of bosons or fermions in traps. In this
case boundary conditions are not a problem since the confinement is directly
insured by the external potential and the boundaries can be rejected to
infinity as is the case of self bound systems.

However, the external potential directly appears in the system Hamiltonian
and thus it is clear that the statistical properties of the system directly
depend upon the considered potential. For example, systems are often
trapped in a one body harmonic potential well $\hat{U}=\sum_{n}\hat{u}_{n}$
where the sum runs over all the particles $n$ and where $\hat{u}=k\hat{r}
^{2}/2.$ 
The corresponding canonical ensemble reads 
\begin{equation}
\hat{D}_{\beta ,k}=\frac{1}{Z_{\beta ,k}}\exp -\beta (\hat{H}+k\hat{R}%
^{2}/2),  \label{EQ:D-pot-ext}
\end{equation}
where $\hat{R}^{2}=\sum_{n}\hat{r}_{n}^{2}$ is the total radius squared.

We can see that the confinement via a potential is equivalent to an extra
constraint defining the compactness of the system $\hat{R}^{2}$
through the introduction of a Lagrange multiplier given by the oscillator
strength.\footnote{Writing eq.(\protect\ref{EQ:D-pot-ext}) we have implicitly
assumed that the confining potential is known with an infinite
accuracy. For realistic physical applications, the external
potential should also be treated using information theory\cite{balian},
and the statistical density matrix should
be averaged over the distribution of $k$ which minimizes the
information (see appendix B).}

\subsubsection{Systems in a fictitious box}

\smallskip 
%Indeed, a consistent statistical description of a finite 
%system would need a precise knowledge of the 
%boundaries, leading to as many thermodynamics 
%as possible boundary conditions. Moreover, in a 
%$D$-dimension space, a boundary condition 
%corresponds to the exact knowledge of the system
%configuration in a $D-1$ dimensional manifold, i.e. 
%in an infinite number of points if the system dimension 
%$D$ is greater than $1$. Such an infinite information is 
%hardly compatible with the very principles of statistical 
%mechanics, as a reduction of our knoledge to a small 
%set of relevant variables. 

In the absence of a physical constraining potential,
systems presenting states in the continuum are often
confined through boundary conditions: a fictitious
container is introduced in the theoretical description of 
unbound systems to limit the Hilbert space 
in order to be able to define 
a statistical ensemble \cite{bondorf}. 
%Those boundaries may correspond to a physical 
%box containig the particles or might be a fictitious 
%box introduced to define a statistical ensemble 
%\QCITE{cite}{}{bondorf}, even if it hardly exists in 
%experimental situations dealing with open systems.
The introduction of such an unphysical box not only
directly affects the energy spectrum and thus
the thermodynamics properties of the system, but also
leads to an intrinsic inconsistency of the statistical theory,
as we now show. 

%Boundaries may account for
%an actual physical container, but more often they are a fictitious
%box introduced in theoretical description of unconfined systems in order to be
%able to define a statistical ensemble \cite{bondorf}. We will discuss in the
%next section how to develop a statistical description of a finite system
%avoiding the introduction of such an unphysical box, but let us first show
%that the boundary conditions lead 
%to an intrinsic inconsistency of the statistical theory, unless 
%properly treated within information theory.

Let us write the eigenvalue problem for an
Hamiltonian operator $\hat{H}$ containing a non-local term such as the
kinetic energy. To be solved we must specify boundary conditions which
directly modify the non-local operators. As an example, let us consider the
standard case %typically with a kinetic term plus some coordinate 
%dependent interaction, 
of the annulation of the wavefunction on the surface $S$ of a containing box 
$V.$ Introducing the projector, $\hat{P}_{S}$ , over the surface $S$ and its
exterior, the boundary conditions reads $\hat{P}_{S}\left| \Psi
\right\rangle =0$ or, using $\hat{P}_{S}^{2}=\hat{P}_{S}$, $\left\langle
\Psi \right| \hat{P}_{S}\left| \Psi \right\rangle =0$. The eigenvalue
equation is then equivalent to a variational principle where the boundary
and the normalization constraints have to be taken into account through the
introduction of two Lagrange multipliers $b_S$ and $E$ 
\begin{equation}
\delta \left( \left\langle \Psi \right| \hat{H}\left| \Psi \right\rangle
-E(\left\langle \Psi \right. \left| \Psi \right\rangle -1)-b_S \left\langle
\Psi \right| \hat{P}_{S}\left| \Psi \right\rangle \right) =0.  \label{modif}
\end{equation}
leading to a modified Schr\"{o}dinger equation 
\begin{equation}
\hat{H}\left| \Psi _{ES}\right\rangle -b_S\hat{P}_{S}\left| \Psi
_{ES}\right\rangle =E\left| \Psi _{ES}\right\rangle  \label{modif2}
\end{equation}
%The two Lagrange multipliers respecting the constraints $\left\langle \Psi
%_{EbS}\right| \hat{P}_{S}\left| \Psi _{EbS}\right\rangle =0$ and $%
%\left\langle \Psi _{EbS}\right. \left| \Psi _{EbS}\right\rangle =1$ are thus 
%$b_{S}$ and $E_{S}$. 
Equation (\ref{modif2}) shows that the energy $%
E $ and wavefunction $\left| \Psi _{ES}\right\rangle $ directly
depend upon the boundary conditions.

Let us now extend this discussion to mixed states. The
boundary condition $\hat{P}_{S}\left| \Psi ^{\left( n\right) }\right\rangle
=0$ is exactly equivalent to the extra constraint 
$<\hat{P}_{S}>=\mathrm{Tr}\hat{D}\hat{P}_{S}=0$.
If we note again $\vec{\hat{A}}$ the observables characterizing 
a given equilibrium, the density matrix including the boundary condition 
reads
%
%Thus, in addition to the observables $\hat{A}_{\ell }$
%constraining the equilibrium one should introduce also $\hat{P}_{S}$ and an
%additional Lagrange multiplier $b.$ Any equilibrium thus reads 
%
\begin{equation}
\hat{D}_{\vec{\lambda }S}=\frac{1}{Z_{\vec{\lambda }S}}\exp -\vec{%
\ \lambda} .\vec{\hat{A}}-b_S\hat{P}_{S}  \label{modif-D-2}
\end{equation}
which shows that the thermodynamics of the system 
does not only depend on the Lagrange multiplier $b_S$, but on the whole
surface $S$. 

%For example, a usual method to deal with open systems 
%not yet expended to infinity is to introduce a fictitious 
%box to define a statistical ensemble \QCITE{cite}{}{bondorf}
%even if it hardly exists in experimental situations. 
%The above demonstration shows that for the very same 
%average particle density we will have as many different 
%thermodynamics as boundary conditions: the knowledge 
%of the particle density is not equivalent to impose a specific 
%constraining box.

For the very same global features such as the
same average particle density or energy, we will have as many different
thermodynamics as boundary conditions.
More important, to
specify the density matrix, the projector $\hat{P}_{S}$ has to be exactly
known and this is in fact impossible. The nature of $\hat{P}_{S}$ is
intrinsically different from the usual global observables $\hat{A}_{\ell }$.
At variance with the $\hat{A}_{\ell }$, $\hat{P}_{S}$ is a many-body operator 
which does not correspond to any physical measurable observable.
The knowledge of $\hat{P}_{S}$ requires the
exact knowledge of each point of the boundary surface while no or few
parameters are sufficient to define the $\hat{A}_{\ell }.$
This infinity of
points corresponds to an infinite amount of information to be known to
define the density matrix (\ref{modif-D-2}). This requirement is in
contradiction with the statistical mechanics principle of minimum
information. Thus eq.(\ref{modif-D-2}) is unphysical.

%Like in the case of the external potential, 
This incoherence should be solved
by a statistical treatment of our knowledge on the boundary condition. 
The density matrix should be an average over the boundary surfaces of 
$\hat{D}_{\vec{\lambda }S}$ weighted by the associated boundary probability
derived applying information theory on our actual knowledge of the
boundaries. A simple way to perform this task is presented in the next
section.

\smallskip  

\subsubsection{Incomplete knowledge on the boundaries}

\smallskip \label{Ch:open} 
%Indeed, a consistent statistical description of a finite 
%system would need a precise knowledge of the 
%boundaries, leading to as many thermodynamics 
%as possible boundary conditions. Moreover, in a 
%$D$-dimension space, a boundary condition 
%corresponds to the exact knowledge of the system
%configuration in a $D-1$ dimensional manifold, i.e. 
%in an infinite number of points if the system dimension 
%$D$ is greater than $1$. Such an infinite information is 
%hardly compatible with the very principles of statistical 
%mechanics, as a reduction of our knoledge to a small 
%set of relevant variables. 

\smallskip One way to get around the difficulties encountered to take into
account our incomplete knowledge on the boundaries  
is to introduce a hierarchy of observables describing the size and
shape of the matter distribution.

%This point of view has the advantage to be also applicable to unconstrained bound 
%or unbound systems. A bound system can always be characterized by a finite 
%size and shape. 
If the system is unbound, the unconstrained
equilibrium corresponds to an infinitely expanded system. 
However, the relevant 
physical problem in general concerns the state of the system  
before this asymptotic stage, when it has only expanded
to a finite volume. This transient stage is constrained by a finite
size, i.e. a finite average of an observable related to the system's size.

For example, if only the average system size $<\hat{R}^{2}>$ is known,
information theory requires the introduction of one additional Lagrange
multiplier imposing this information, i.e. one of the $\hat{A}_{\ell }$
describing the system is $\hat{R}^{2}$.
\footnote{
Depending upon the actual system studied, more observables describing its
size or shape might be needed. Then, all those observables should be introduced
as constraints in the minimization of the information in order to obtain an
adequate statistical description.}
If the additional information reduces to the
energy, the Lagrange multipliers associated with the state variables 
$<\hat{H}>$ and $<\hat{R}^{2}>$ are respectively $\lambda _{H}=\beta =1/T$, 
the inverse of a temperature and $\lambda _{R^{2}}$, which has the 
dimension of a pressure when divided by a typical scale $R_{0}$ 
and by the temperature, $\lambda _{R^{2}}=\beta PR_{0}$. 
The minimum information principle implies
\begin{equation}
\hat{D}_{\beta ,P}=\frac{1}{Z_{_{\beta ,P}}}\exp -\beta \left( \hat{H}+PR_{0}%
\hat{R}^{2}\right) ,  \label{EQ:D-T-P}
\end{equation}
which is akin to an isobar canonical ensemble. It is interesting to remark that
the term proportional to $\hat{R}^{2}$ can be equivalently viewed as a mean
square radius constraint or as an harmonic external potential modifying the
effective Hamiltonian of the system as in the equilibrium (\ref{EQ:D-pot-ext}
). Similarly the expectation value $<\hat{H}+PR_{0}\hat{R}^{2}> $ can be
interpreted as an enthalpy or as a constrained energy.

%\subsubsection{\protect\smallskip An example: the freeze-out configuration}
%\smallskip  
%Many physical systems experimentally accessible are neither trapped in a
%physical external potential nor confined in an box. 

A typical application of eq.(\ref{EQ:D-T-P}) is given by the 
freeze-out hypothesis proposed for the unconfined transient
finite systems produced in atomic or nuclear
collisions\cite{high:energy,low:energy}. At a given time the main
evolution (i.e. the main creation of entropy) is assumed to stop. Before
this freeze-out time the evolution is assumed to be complex enough such that
all the partitions compatible with common gross features are freely (i.e.
with no bias) explored. After the freeze-out time, partitions are supposed
to be essentially frozen because of the lack of interactions. Typically
thermal and chemical equilibrium is assumed, meaning that the information on
the energetics and particle numbers is limited to the observables $<\hat{H}>$
and $<\hat{N}_{f}>$ for the different species $f$ \cite{high:energy,bondorf}.

The freeze-out occurs when the system has expanded to a finite size.
Then at least one measure of the system's compactness
should be included among the collective variables characterizing the 
statistical ensemble. The limited knowledge of the system extension
leads to a minimum biased density matrix given by eq.(\ref{EQ:D-T-P}).
\footnote{%
The ensemble of events dynamically prepared may be characterized by an
information about size and shape more complex than the simple mean square
root radius. In this case other constraints can be introduced such as $\hat{R%
}^{4}$ if the radii fluctuations are not maximal (i.e. are not maximizing
the entropy, meaning that fluctuations are containing non-trivial
information ) or $\hat{Q }_{2}=2\hat{Z}^{2}-\left( \hat{X}^{2}+\hat{Y}
^{2}\right) $ if the system is not spherical in average but has a finite
quadrupole deformation.}
%This can also be seen as a consequence of the finite time needed to reach the
%freeze-out configuration.  

Such an ensemble does not suffer
from the drawbacks of models introducing a constraining box: i.e. both the
arbitrariness of such a fictitious box and the conceptual problems of the
statistical treatment of boundary conditions.

\smallskip 
To conclude this section, we have shown that all the different ways to 
describe boundary conditions in finite unbound systems 
lead to the introduction of additional constraints in the set of
relevant observables $\hat{A}_{\ell }$ describing the statistical ensemble.
The more information are known on the spatial extension of the system, 
the more complex are the additional observables. In other words,  
only Gibbs equilibria with compactness constraints are physically meaningful
when discussing small systems in the presence of continuum states. 
Since outside the thermodynamic limit
statistical ensemble are never equivalent, the thermodynamics of finite
unbound systems always depends on the compactness conditions.

%\subsection{\protect\smallskip The dynamics of the expansion}

%Let us now apply the above formalism to physically important situations of
%transient unconfined  systems.

%The standard Gibbs equilibrium eq. (\ref{EQ:D}) with one of the observable
%related to the system extension such as the example (\ref{EQ:D-T-P}), is
%the simplest statistical description of a finite unconfined system at the
%time of entropy saturation. \footnote{%
%This expression can be generalized to the case of observables freezing at
%different times through the introduction of time odd observables that take
%into account the dynamical evolution of the system between the different
%freeze out times, see section \ref{multiple}.} In the absence of a recalling
%force, as time goes on the system size will increase. In the next sections 
%\ref{ideal-1},\ref{ideal-2} we shall describe this situation in the
%simplified case of an ideal system initially confined in an arbitrary
%density state, and successively expanding freely in the vacuum. We will show
%that the out of equilibrium expansion can be exactly accounted for at any
%time within a finite amount of information, with the natural emergence of
%time odd collective flow observables among the state variables. The
%situation is then generalized to the inclusion of a one body interaction
%(section \ref{potential}).

\subsection{Unconfined finite ideal gas}

\label{ideal-1}

\smallskip Let us first consider the case of a finite ideal gas, i.e. the
Hamiltonian is reduced to the one body kinetic term 
\begin{equation}
\hat{H}=\hat{K}\equiv \sum_{n}\frac{\vec{\hat{p}}_{n}^{2}}{2m}\equiv 
\frac{\vec{\hat{P}}^{2}}{2m}.  \label{EQ:H-osc}
\end{equation}

Let us assume that the first constraining observable is $\hat{A}_{1}=\hat{K}$
and so the associated Lagrange multiplier is the usual inverse temperature $%
\beta $. As discussed in subsection \ref{boundary}, 
any statistical ensemble is ill-defined unless boundary conditions are 
specified. In the case of an unbound system, one is forced to 
introduce additional observables constraining the spatial extension 
of the system to a finite size. 
This is a situation often encountered experimentally: 
a finite system of loosely interacting particles or clusters 
with a finite extension 
after an expansion during a limited time 
in an open space. 
The minimal assumption needed to characterize statistically
the ensemble of states, is given by one single observable related 
to the size, e.g. the knowledge at a given time of the mean
square radius $<\vec{\hat{R}}^{2}>$ (with $\vec{\hat{R}}^{2}=\sum_{n}
\vec{\hat{r}}_{n}^{2}$).
Then we have to introduce the constraining observable $\hat{A}_{2}=\vec{
\hat{R}}^{2}$ associated with a Lagrange multiplier $\lambda _{0}$
in the statistical description.  
The maximum entropy solution is given by 
%\begin{equation}
%\hat{D}_{\beta \lambda _{0}}=\frac{1}{Z_{_{\beta \lambda _{0}}}}\exp -\beta 
%\hat{H}-\lambda _{0}\vec{\hat{R}}^{2}
%\end{equation}
%Rewriting the above distribution as 
\begin{equation}
\hat{D}_{\beta \lambda _{0}}=\frac{1}{Z_{_{\beta \lambda _{0}}}}\exp -\beta
\sum_{n}\left( \frac{\vec{\hat{p}}_{n}^{2}}{2m}+\frac{\lambda _{0}}{\beta 
}\vec{\hat{r}}_{n}^{2}\right) .  \label{EQ:D-osc}
\end{equation}
%It is interesting to see that up to now we have not used the 
%expression of the Hamiltonian. 
Eq.(\ref{EQ:D-osc}) 
%shows that, independent of the interparticle interactions, 
%the minimum biased distribution for a finite size unconfined 
%system with a finite mean square radius,
is akin to a system of non-interacting particles trapped in an harmonic
oscillator potential with a string constant $k=2\lambda _{0}/\beta .$ From
the partition sum, the EOS are easily derived. For example, in the classical
case using the equipartition and the virial theorem it is easy to derive the
system EOS for each particle $n=1,\dots ,N$ 
\[
\frac{<\vec{\hat{p}}_{n}^{2}>}{2m}=\frac{\lambda _{0}}{\beta }<\vec{%
\hat{r}}_{n}^{2}>=\frac{3}{2\beta }, 
\]
i.e. 
\[
<\vec{\hat{p}}_{n}^{2}>=\frac{3m}{\beta }\;\;\;;\;\;\;<\vec{\hat{r}}
_{n}^{2}>=\frac{3}{2\lambda _{0}}. 
\]
Since in the presented derivation, the $\lambda _{0}\vec{\hat{R}}^{2}$
term is not an external confining potential but only a finite size
constraint 
%defined 
%at a specific time $t_{0}$ and taken into account through the 
%use of information theory
, the minimum biased distribution (\ref{EQ:D-osc}) is not stationary 
($\{\hat{H},\hat{D}_{\beta \lambda _{0}}\}\neq 0$). 
The system represented by eq.(\ref{EQ:D-osc}) at a time $t_{0}$ 
will evolve in time according to the Hamiltonian (\ref{EQ:H-osc}).

The physical situation we are describing can be explicitly realized
experimentally by taking a system in equilibrium in an harmonic trapping
potential $k\vec{\hat{r}}_{n}^{2}/2$ and by suddenly removing the
confining potential at a time $t_0$. 
After that, the non-interacting particles freely expand
ballistically (see section \ref{harmonic}). In this case the information
theory ansatz (\ref{EQ:D-osc}) with $\lambda _{0}=k\beta /2$  
describes the standard static equilibrium just before the trapping
potential is removed, and is therefore the exact initial state. 
This is akin to the actual experimental procedure used
in Bose condensates studies\cite{bose,traps}. This physical situation is
also the picture underlying the freeze-out hypothesis used to describe data
in heavy ion collisions (see section \ref{Ch:open}). 
At a given time the system is supposed to occupy a
given volume and to behave thereafter like an ideal gas of fragments or
particles. At variance with experiments in traps, for collision experiments
the minimum biased distribution (\ref{EQ:D-osc}) at time $t_{0}$ is only an
ansatz, implying that other observables might be needed to correctly
describe the density matrix at the initial time $t_{0}$. This more
complicated case is studied in the next section.

What is common to both these situations, is that the state (\ref{EQ:D-osc})
will evolve in time. So let us apply the formalism developed in section \ref
{multiple} assuming that $<\vec{\hat{R}}^{2}>$ is known at a time $t_{0}$
while the system is observed at a later time $t.$ Since energy is a trivial
constant of any Hamiltonian motion, $\{\hat{H},\hat{H}\}=0,$ we do not have
to introduce a specific time for the observation of the energy.

According to section (\ref{multiple}) to take into account the time
evolution, we must introduce additional constraining observables 
\begin{eqnarray*}
\hat{B}_{R}^{(1)} &=&\{\hat{H},\vec{\hat{R}}^{2}\}=-\sum_{n}\frac{1}{m}%
\left( \vec{\hat{p}}_{n}\cdot \vec{\hat{r}}_{n}+\vec{\hat{r}}%
_{n}\cdot \vec{\hat{p}}_{n}\right) \\
\hat{B}_{R}^{(2)} &=&\{\hat{H},\hat{B}_{R}^{(1)}\}=\sum_{n}\frac{2\vec{%
\hat{p}}_{n}^{2}}{m^{2}}.
\end{eqnarray*}
Since $\{\hat{H},\hat{B}_{R}^{(2)}\}=0$, all the other $\hat{B}_{R}^{(p)}$
with $p>2$ are zero. The above relations are valid both in quantum and
classical mechanics. Then the most general density matrix corresponding
to the time-dependent Gibbs ensemble is given by
\begin{equation}
\hat{D}_{\beta ,\lambda _{0}}(t)=\frac{1}{Z_{\beta ,\lambda _{0}}}\exp
\sum_{n}-\beta ^{\prime }\left( t\right) \frac{\vec{\hat{p}}_{n}^{2}}{2m}
-\lambda _{0}\vec{\hat{r}}_{n}^{2}+\frac{\nu _{0}\left( t\right) }{2}
\left( \vec{\hat{p}}_{n}\cdot \vec{\hat{r}}_{n}+\vec{\hat{r}}%
_{n}\cdot \vec{\hat{p}}_{n}\right) ,  \label{EQ:GP-expan}
\end{equation}
with 
\begin{equation}
\beta ^{\prime }\left( t\right) =\beta +2\lambda _{0}\left( t-t_{0}\right)
^{2}/m  \label{EQ:beta-t-}
\end{equation}
and 
\begin{equation}
\nu _{0}\left( t\right) =2\lambda _{0}\left( t-t_{0}\right) /m.
\label{EQ:nu-t-}
\end{equation}
Let us explicitly show that the density matrix (\ref{EQ:GP-expan}) can be
interpreted as a radially expanding ideal gas. Indeed the distribution can
be written as 
\begin{equation}
\hat{D}_{\beta ,\lambda _{0}}(t)=\frac{1}{Z_{\beta ,\lambda _{0}}}\exp
\sum_{n}-\beta ^{\prime }\left( t\right) \frac{\left( \vec{\hat{p}}
_{n}-mh_{0}\left( t\right) \vec{\hat{r}}_{n}\right) ^{2}}{2m}-\lambda
_{0}^{\prime }\left( t\right) \vec{\hat{r}}_{n}^{2}  \label{EQ:D-expan}
\end{equation}
where the Hubblian factor reads 
\begin{equation}
h_{0}=\frac{\nu _{0}\left( t\right) }{\beta ^{\prime }\left( t\right) }=%
\frac{2\lambda _{0}\left( t-t_{0}\right) }{\beta m+2\lambda \left(
t-t_{0}\right) ^{2}}  \label{EQ:h0}
\end{equation}
while the confining Lagrange multiplier is transformed into 
\begin{equation}
\lambda ^{\prime }\left( t\right) =\lambda _{0}-\frac{m}{2}\frac{\nu
_{0}^{2}\left( t\right) }{\beta ^{\prime }\left( t\right) }=\frac{\lambda
_{0}\beta m}{\beta m+2\lambda _{0}\left( t-t_{0}\right) ^{2}}
\label{EQ:lambda-prime}
\end{equation}
In the density matrix (\ref{EQ:D-expan}) the term $mh_{0}\left( t\right) 
\vec{\hat{r}}_{n}$ correcting the momentum can be interpreted as a
collective motion produced by a radial velocity $h_{0}\left( t\right) \vec{\hat{r}}
_{n}$. This proportionality of the velocity with $\hat{r}_{n}$ shows that
the motion is akin to a  self-similar Hubble expansion. As a
consequence, when this collective motion is subtracted from the particle
momentum, the density matrix (\ref{EQ:D-expan}) corresponds at any time to a
standard equilibrium (\ref{EQ:D-osc}) in the local rest frame. 
%Therefore, the density matrix (\ref{EQ:GP-expan}) corresponds
%to an ideal gas locally at equilibrium and following a collective
%self-similar expansion.

Computing the time evolution of the distribution (\ref{EQ:D-expan}) under
the action of the Hamiltonian (\ref{EQ:H-osc}), it is easy to verify that
the information theory ansatz (\ref{EQ:D-expan}) is, at every time $t$, the
exact solution of an ideal gas initially trapped in an harmonic oscillator
potential up to the time $t_{0}$, and then freely expanding in the vacuum
after the confining potential has been suppressed. In this case the infinite
information which is a priori needed to follow the large-time evolution of the
density matrix according to eq.(\ref{multistep}), reduces to the three
observables $\vec{\hat{r}}^{2}$, $\vec{\hat{p}}^{2}$, $\vec{\hat{r}}
\cdot \vec{\hat{p}}+\vec{\hat{p}}\cdot \vec{\hat{r}}$. Indeed these
operators form a closed Lie algebra containing the Hamiltonian 
operator, implying that 
%all the other $\vec{
%\hat{B}}^{(p)}$ operators are identically zero and that 
the exact 
evolution of (\ref{EQ:D-expan}) preserves it algebraic structure.

An easy way to follow this time dependence is to compute the evolution of
the 3 averages $<\vec{\hat{r}}_{n}^{2}>$ , $<\vec{\hat{p}}_{n}\cdot 
\vec{\hat{r}}_{n}+\vec{\hat{r}}_{n}\cdot \vec{\hat{p}}_{n}>$ and $<
\vec{\hat{p}}_{n}^{2}>.$ Using $\partial _{t}<\hat{A}>=-\{\hat{H},\hat{A}
\}$ we can easily show that 
\begin{eqnarray*}
&&\partial _{t}<\vec{\hat{r}}_{n}^{2}>=\frac{1}{m}<\vec{\hat{p}}%
_{n}\cdot \vec{\hat{r}}_{n}+\vec{\hat{r}}_{n}\cdot \vec{\hat{p}}%
_{n}> \\
&&\partial _{t}<\vec{\hat{p}}_{n}\cdot \vec{\hat{r}}_{n}+\vec{\hat{r%
}}_{n}\cdot \vec{\hat{p}}_{n}>=\frac{2}{m}<\vec{\hat{p}}_{n}^{2}>.
\end{eqnarray*}
The energy being constant, $<\vec{\hat{p}}_{n}^{2}>$ is also constant so
that the radial momentum $<\vec{\hat{p}}_{n}\cdot \vec{\hat{r}}_{n}+
\vec{\hat{r}}_{n}\cdot \vec{\hat{p}}_{n}>$ linearly depends upon time
while the motion of $<\vec{\hat{r}}_{n}^{2}>$ is uniformly accelerated.
This evolution of the constraints corresponds the dynamics of $\beta ^{\prime
} $, $\nu _{0}$ and $\lambda ^{\prime }$ deduced above.

The description of the time evolution when considering unconfined finite
systems has introduced a new phenomenon: the expansion. One should then
consider a more general equilibrium of a finite-size expanding
finite-systems with $\beta ^{\prime }$, $h_{0}$ and $\lambda _{0}^{\prime }$
as free parameters. Then, if the observed minimum biased distribution at time $%
t$ is coming from a confined system at time $t_{0}$, the three parameters $
\beta ^{\prime }$, $h_{0}$ and $\lambda _{0}^{\prime }$ should be linked to
the time $t_{0}$, the initial temperature $\beta ^{-1}$ and the initial $
\lambda _{0}$ by equations (\ref{EQ:beta-t-}), (\ref{EQ:h0}) and (\ref
{EQ:lambda-prime})\cite{npa}.

The important consequence of that is that radial flow is a necessary
ingredient of any statistical description of unconfined finite systems
in the presence of a continuum: the
static (canonical or microcanonical) Gibbs ansatz in a confining box which
is often employed\cite{bondorf} misses this crucial point. On the other
hand, if a radial flow is observed in the experimental data, the formalism
we have developed allows to associate this flow observation to a
distribution at a former time when flow was absent. This initial
distribution corresponds to a standard static Gibbs equilibrium in a
confining harmonic potential, i.e. to an isobar ensemble. 
We will see in the following that more complex non-Hubblian
flows can be associated with additional constraints on the matter distribution. 

\subsection{\protect\smallskip Deformed unconfined systems}

\label{ideal-2} \label{deformed}

The systems formed in many experimental situations, ranging from asymmetric
atomic traps\cite{traps} to non-central or incompletely damped 
nuclear collisions\cite{low:energy},
present an explicit 
%spatial 
deformation. The simplest one is of quadrupolar
nature, and corresponds to a non-zero $<\hat{Q}_{20}>$ where $\hat{Q}_{20}$
is the standard quadrupole moment 
$\hat{Q}_{20}
%=2\hat{Z}^{2}-(\hat{X}^{2}+\hat{Y}^{2})
=\sum_{n}\hat{q} _{20_{n}}$ 
where the single particle operator
reads $\hat{q}_{20_{n}}=2\hat{z }_{n}^{2}-(\hat{x}_{n}^{2}+\hat{y}_{n}^{2})$
. Therefore, the constraining observable $\hat{Q}_{20}$ associated with the
Lagrange multiplier $\lambda _{2}$ should be introduced in the definition of
the out of  (spherical) equilibrium statistical ensemble. In addition, 
following the procedure
explained above to take into account the 
influence of the time dependence, 
this observable should be complemented by its multiple
commutators with $\hat{H}$ 
\begin{equation}
\hat{B}_{Q}^{(1)} =\{\hat{H},\hat{Q}_{20}\} =-\frac{1}{m}\sum_{n}2\hat{p}
_{z_{n}}\hat{z}_{n}-\hat{p} _{x_{n}}\hat{x}_{n}-\hat{p}_{y_{n}}\hat{y}_{n}+(
\vec{p}\leftrightarrow \vec{r})
\end{equation}
which is nothing but an asymmetric flow, and

\begin{equation}
\hat{B}_{Q}^{(2)} =\{\hat{H},\hat{B}_{Q}^{(1)}\} =\frac{2}{m^{2}}\sum_{n}2%
\hat{p}_{z_{n}}^{2}-\hat{p} _{x_{n}}^{2}-\hat{p}_{y_{n}}^{2},
\end{equation}
a deformation in $p$ space. Again, since $\{\hat{H},\hat{B}_{Q}^{(2)}\}=0$,
all the other $\hat{B}_{Q}^{(p)}$ with $p>2$ are zero.

Then the minimum biased density matrix characterized at a time $t_{0}$ by
the compactness $<\vec{\hat{R}}^{2}>$ associated with a Lagrange
multiplier $\lambda _{0}$, a quadrupole deformation $<\hat{Q}_{20}>$ at a
time $t_{2}$ imposed by a second Lagrange multiplier $\lambda _{2}$, and by
an average energy $<\vec{\hat{P}}^{2}/2m>$ 
%taken into account through the usual introduction of a
%temperature $\beta ^{-1}$
, is given by
\begin{eqnarray}
\hat{D}_{\beta ,\lambda _{0},\lambda _{2}} &(t)=&\frac{1}{Z_{\beta ,\lambda
_{0},\lambda _{2}}}\exp \sum_{n}-\beta _{z}^{\prime }\left( t\right) \frac{%
\hat{p}_{z_{n}}^{2}}{2m}-\lambda _{z}\hat{z}_{n}^{2}+\frac{\nu _{z}\left(
t\right) }{2}\left( \hat{p}_{z_{n}}\hat{z}_{n}\vec{+}\hat{z}_{n}\hat{p}
_{z_{n}}\right) \cdot  \nonumber \\
&\exp &\sum_{n}-\beta _{\perp }^{\prime }\left( t\right) \frac{\vec{\hat{p%
}}_{\perp _{n}}^{2}}{2m}-\lambda _{\perp }\vec{\hat{r}}_{\perp _{n}}^{2}+%
\frac{\nu _{\perp }\left( t\right) }{2}\left( \vec{\hat{p}}_{\perp
_{n}}\cdot \vec{\hat{r}}_{\perp _{n}}+\vec{\hat{r}}_{\perp _{n}}\cdot 
\vec{\hat{p}}_{\perp _{n}}\right) ,  \label{EQ:deform-equil}
\end{eqnarray}
where we have introduced the momentum ($\vec{\hat{p}}_{\perp _{n}}$) and
coordinate ($\vec{\hat{r}}_{\perp _{n}}$) perpendicular to the
deformation axis $z.$ The distribution at a time $t$ (\ref{EQ:deform-equil})
is then function of the spatial deformation parameters 
$\lambda _{z} =\lambda _{0}+2\lambda _{2}$ and
$\lambda _{\perp } =\lambda _{0}-\lambda _{2}$.
%\begin{eqnarray}
%\lambda _{z} &=&\lambda _{0}+2\lambda _{2}, \\
%\lambda _{\perp } &=&\lambda _{0}-\lambda _{2}.
%\end{eqnarray}
We can see that the kinetic energy is not determined by a unique temperature
but by two different Lagrange parameters 
\begin{eqnarray}
\beta _{z}^{\prime }\left( t\right) &=&\beta +2(\lambda _{0}\left(
t-t_{0}\right) ^{2}+2\lambda _{2}\left( t-t_{2}\right) ^{2})/m \\
\beta _{\perp }^{\prime }\left( t\right) &=&\beta +2(\lambda _{0}\left(
t-t_{0}\right) ^{2}-\lambda _{2}\left( t-t_{2}\right) ^{2})/m
\end{eqnarray}
Finally the deformed expansion parameters read 
\begin{eqnarray}
\nu _{z}\left( t\right) &=&2(\lambda _{0}\left( t-t_{0}\right) +2\lambda
_{2}\left( t-t_{2}\right) )/m \\
\nu _{\perp }\left( t\right) &=&2(\lambda _{0}\left( t-t_{0}\right) -\lambda
_{2}\left( t-t_{2}\right) )/m.
\end{eqnarray}
The minimum biased density matrix (\ref{EQ:deform-equil}) can also be
written as a generalized equilibrium in the local rest frame 
\begin{eqnarray}
\hat{D}_{\beta ,\lambda _{0},\lambda _{2}}(t) &=&\frac{1}{Z_{\beta ,\lambda
_{0},\lambda _{2}}}\exp \sum_{n}-\beta _{z}^{\prime }\left( t\right) \frac{%
\left( \hat{p}_{z_{n}}-mh_{z}\left( t\right) \hat{z}_{n}\right) ^{2}}{2m}
-\lambda _{z}^{\prime }\hat{z}_{n}^{2}  \label{EQ:deform-equil2} \\
&&\exp \sum_{n}-\beta _{\perp }^{\prime }\left( t\right) \frac{\left( 
\vec{\hat{p}}_{\perp _{n}}-mh_{\perp }\left( t\right) \vec{\hat{r}}
_{\perp _{n}}\right) ^{2}}{2m}-\lambda _{\perp }^{\prime }\vec{\hat{r}}
_{\perp _{n}}^{2}
\end{eqnarray}
where the Hubblian flow factors read 
\begin{equation}
h_{z}=\frac{\nu _{z}\left( t\right) }{\beta _{z}^{\prime }\left( t\right) }
;\;\;\;h_{\perp }=\frac{\nu _{\perp }\left( t\right) }{\beta _{\perp
}^{\prime }\left( t\right) },
\end{equation}
while the confining Lagrange multipliers are transformed into 
\begin{equation}
\lambda _{z}^{\prime }\left( t\right) =\lambda _{z}-\frac{m}{2}\frac{\nu
_{z}^{2}\left( t\right) }{\beta _{z}^{\prime }\left( t\right) }
;\;\;\;\lambda _{\perp }^{\prime }\left( t\right) =\lambda _{\perp }-\frac{m%
}{2}\frac{\nu _{\perp }^{2}\left( t\right) }{\beta _{\perp }^{\prime }\left(
t\right) }.
\end{equation}

Computing the exact time evolution of the density matrix (\ref
{EQ:deform-equil2}), and using again the fact that for all $k$,
representing both the particle and the axis labels, $\hat{p}
_{k}^{2},$ $\hat{r}_{k}^{2}$ and $(\hat{p}_{k}\hat{r}_{k}+\hat{r}_{k}\hat{p}
_{k})$ %,where $\hat{r}_{k}$
%and $\hat{p}_{k}$ are the $k$ 's component of the $\vec{\hat{r}}$ and $%
%\vec{\hat{p}}$ vectors, 
form a closed Lie algebra, it is easy to demonstrate that the density matrix
(\ref{EQ:deform-equil2}) is, at every time $t$, the exact solution of 
the dynamical evolution. For example, an
ideal gas initially trapped in a quadrupoly deformed harmonic oscillator
potential up to the time $t_{0}=t_{2}$, and then freely expanding in the vacuum up
to the time $t$ exactly follows (\ref{EQ:deform-equil2}). 
The time $t_{2}$ start to play a role 
if the trapping potentials in different directions are not switched-off
 at the same time. For such a case the density  (\ref{EQ:deform-equil2})
 remains the exact solution.

It is interesting to note that the $\lambda $'s are inversely proportional
to the size of the system in coordinate space, and so are the produced
flows. Thus the deformation 
%($h$) 
in $p$ space is larger in the narrower
direction in the coordinate space.

Taking advantage of the 
commutator algebra of various functionals of the operators $
\vec{\hat{p}}$ and $\vec{\hat{r}}$ this discussion can be
generalized to more complex multipolar deformations and flows (see
appendix A).

\subsection{Role of an external potential}

\label{potential} 
 
In many physical situations the Hamiltonian of the system does not reduce to
a simple ideal gas, but also may contain explicitly a potential term $\hat{U}
=\sum_{n}U\left( \vec{\hat{r}}_{n}\right) $, either coming from the
interaction with some external field, or from a self interaction treated at
the mean-field level. In such a case the Hamiltonian becomes $\hat{H}=$ $%
\sum_{n}\vec{\hat{p}}_{n}^{2}/2m+U\left( \vec{\hat{r}}_{n}\right) .$
Considering the additional constraint $\vec{\hat{R}}^{2}$ it is easy to
show that $\hat{B}_{R}^{(1)}=-\sum_{n}\frac{1}{m}\left( \vec{\hat{ p}}
_{n}\cdot \vec{\hat{r}}_{n}+\vec{\hat{r}}_{n}\cdot \vec{\hat{p}}
_{n}\right) $ remains unchanged since $\left\{ U\left( \vec{\hat{r}}
_{n}\right) ,\vec{\hat{r}}_{n}^{2}\right\} =0.$ The difference starts at
the second order, since in addition to the term $\sum_{n}2\vec{\hat{p}}
_{n}^{2}/m^{2}$ , $\hat{B}_{R}^{(2)}$ contains 
\[
\{\hat{U},\hat{B}_{R}^{(1)}\}
%=-\sum_{n}\left\{ U\left( \vec{\hat{r}}
%_{n}\right) ,\frac{1}{m}\left( \vec{\hat{p}}_{n}\cdot \vec{\hat{r}}
%_{n}+\vec{\hat{r}}_{n}\cdot \vec{\hat{p}}_{n}\right) \right\}
 =-
\sum_{n}\frac{2\vec{\hat{r}}_{n}\cdot \vec{\nabla }U\left( \vec{%
\hat{r}} _{n}\right) }{m} . 
\]
%It should be noticed that if $U\left( \vec{\hat{r}}_{n}\right) =U\left( 
%\hat{r}_{n}=\left| \vec{\hat{r}}_{n}\right| \right) $ then the additional
%term in $\hat{B}_{R}^{(2)}$ reads $\sum_{n}\hat{r}_{n}\partial _{r}U\left( 
%\hat{r}_{n}\right) /m.$ 
For short time evolutions (up to the second order in $\delta t=t-t_0$) the
minimum biased density matrix corresponds to a self similar flow in a
modified potential 
\[
\hat{D}_{\beta ,\lambda _{0}}(t)=\frac{1}{Z_{\beta ,\lambda _{0}}}\exp
\sum_{n}-\beta ^{\prime }\left( t\right)\left ( \frac{\left( \vec{\hat{p}}
_{n}-mh_{0}\left( t\right) \vec{\hat{r}}_{n}\right) ^{2}}{2m}+ \hat{U}
^{\prime} \right ) 
\]
where the time dependent temperature $\beta ^{\prime }$ and the Hubblian
factor $h_0$ are given by eqs.(\ref{EQ:beta-t-}),(\ref{EQ:h0}), and the
effective potential $\hat{U}^{\prime}$ is given by 
\begin{equation}
\hat{U}^{\prime}= \frac{\beta }{\beta^{\prime}(t)} \hat{U}
+\sum_n \frac{\lambda_{0}}{\beta ^{\prime }(t)} \vec{\hat{r}}
_{n}^{2} - \frac{\lambda _{0} \delta t^2 }{m\beta^{\prime}(t)} \left ( \frac{
2\lambda_{0}\vec{\hat{r}}_{n}^{2}} {\beta^{\prime }(t)} + \vec{\hat{r}}
_{n}\cdot \vec{\nabla }U\left( \vec{\hat{r}} _{n}\right) \right )
\end{equation}
In the special case of an harmonic potential $U\left( \vec{\hat{r}}
\right)=k\vec{\hat{r}}^2/2$ the solution can be exactly worked out at any
arbitrary time. Indeed in this case the $\hat{B^{(p)}}$ operators read for $%
p\geq 1$ 
\begin{eqnarray*}
\hat{B}^{(2p)}&=& \sum_n (-1)^p \left ( 2 \omega \right)^{2p} \left ( \frac{%
\vec{\hat{r}}_{n}^{2}}{2}- \frac{\vec{\hat{p}}_{n}^{2}}{2mk}\right ) \\
\hat{B}^{(2p+1)}&=& -\sum_n (-1)^p \left ( 2 \omega \right)^{2p} \frac{%
\vec{\hat{p}}_{n}\cdot \vec{\hat{r}}_{n} + \vec{\hat{r}}_{n} \cdot 
\vec{\hat{p}}_{n} }{m}
\end{eqnarray*}
where $\omega^2=k/m$.
The distribution at time $t$ reads 
\[
\hat{D}_{\beta^{\prime} ,\lambda^{\prime}_0,\nu }(t)= \frac{1}{
Z_{\beta^{\prime} ,\lambda^{\prime}_0,\nu }}\exp \sum_{n}-\beta ^{\prime }(t) 
\frac{\vec{\hat{p}}_{n}^{2}}{2m} -\lambda_{0}^{\prime }\left( t\right) 
\vec{\hat{r}}_{n}^{2} +\frac{\nu(t)}{2
m} \left ( \vec{\hat{p}}_{n}\cdot 
\vec{\hat{r}}_{n} + \vec{\hat{r}}_{n} \cdot \vec{\hat{p}}
_{n}\right
) , 
\]
where the temperature is oscillating in time 
\[
\beta ^{\prime }\left( t\right) =\beta - \frac{\lambda _{0}}{m}
\frac{\left ( \cos
2 \omega\left( t-t_{0}\right)-1\right )}{\omega^2} 
\]
as well as the effective constraining field 
\[
\lambda_0 ^{\prime }\left( t\right) =\beta \frac{m\omega^2}{2} + \lambda_0 \left ( 1 + \frac{1}{2}
\left
(\cos 2 \omega\left( t-t_{0}\right)-1\right )\right ) 
\]
and the collective radial velocity 
\[
\nu \left( t\right) =
%2
\lambda _{0}
%\left( t-t_{0}\right)
\frac {\sin 2  
\omega\left( t-t_{0}\right)} {
%2  
m \omega %\left( t-t_{0}\right)
} . 
\]
Similar to the case of an ideal gas section \ref{ideal-1}, this density
matrix can be interpreted as a Gibbs equilibrium in the rest frame of a
breathing system 
\begin{equation}
\hat{D}_{\beta^{\prime } ,\lambda^{\prime }}(t)= \frac{1}{Z_{\beta^{\prime }
,\lambda^{\prime }}}\exp \sum_{n}-\beta ^{\prime }\left( t\right) \frac{%
\left( \vec{\hat{p}} _{n}-mh^{\prime }\left( t\right) \vec{\hat{r}}
_{n}\right) ^{2}}{2m} -\lambda^{\prime }\left( t\right) \vec{\hat{r}}
_{n}^{2}
\end{equation}
where the Hubblian factor reads 
\begin{equation}
h^{\prime }(t)=
\frac {\lambda _{0}}{\beta^{\prime }(t)}
%\left( t-t_{0}\right)
\frac {\sin 2  
\omega\left( t-t_{0}\right)} {
%2  
m \omega %\left( t-t_{0}\right)
} 
%
%\frac{2\lambda^{\prime }_{0}\left( t-t_{0}\right) }{
%m\beta^{\prime }}
\end{equation}
and the effective pressure is given by 
\begin{equation}
\lambda ^{\prime }\left( t\right) =\lambda^{\prime }_{0}(t)
-
\frac {\lambda _{0}^2}{2\beta^{\prime }(t)}
%\left( t-t_{0}\right)
\frac {\left(\sin 2  
\omega\left( t-t_{0}\right)\right)^2} {
%2  
m \omega^2 %\left( t-t_{0}\right)
} 
%
%- \frac{
%2\lambda^{\prime } _{0}\left( t-t_{0}\right) ^{2}}{m \beta^{\prime}}
\end{equation}

Due to the confining harmonic potential the distribution is periodic in
time. The collective flow is oscillating but it is interesting to note that
the velocity is at all times proportional to the radius. The expansion
deviates from self similarity only in presence of an anharmonic potential or
of finite range two body interactions, as it is shown in appendix B.

\section{Conclusion}

In this paper we have introduced an extension of Gibbs ensembles to 
account for time
dependent constraints, in classical as well as in quantum mechanics. The
formalism is developed for generic equations of motion and is
illustrated on both Hamiltonian and dissipative time evolutions of the
density matrix. We show that the time dependence imposes the introduction
of new constraining observables. In the case of an Hamiltonian dynamics 
theses new relevant informations are the multiple commutators of the 
initial observable with $\hat{H}$. This leads to time odd constraints
which can be interpreted as collective flows.

This formalism gives a statistical description of a system %observed (or
characterized by some relevant observables
%) 
%(i.e. relevant average information are measured) 
defined at a time at which the entropy has not reached its saturating
 value yet, as
it may be the case in intermediate energy heavy ion reactions\cite
{low:energy}. Another physical application concerns systems for which the
relevant observables pertain to different times, as in high energy nuclear
collisions where the kinetic energy seems to be still dissipated when the
chemistry of the system is fixed\cite{high:energy} (i.e. systems presenting
different freeze-out times for different observables).

An important result is that any statistical description of an
unbound 
 finite
system must necessarily contain a local collective velocity term. Indeed the
knowledge of the average spatial extension of the system at a given time,
naturally produces a flow constraint at any successive times. 
This describes the dynamical expansion of the system. 
This is important
for the transient unconfined systems formed in collisional processes
which freely expand in vacuum. 
It may also describe out of equilibrium systems bound 
in a self consistent mean field or 
 confined in traps. 
Conversely a collective flow measurement at a given time can be
translated into an information on the system extension at a former time.

In the general case the time dependence of the density matrix can be
accounted by introducing an infinite set of time even as well as time odd
constraints and the relevant information (i.e. the number of independent
relevant observables or state variables) 
%diverges 
rapidly increases
 with time. In some cases
however the state variables form a closed Lie algebra, and the knowledge of
a finite number of observables at a given time allows to predict the total
density matrix at any successive time. This is notably the case of a system
of particles interacting through polynomial one-body or two-body 
potentials.
% of the type $|\vec{
%\ r}_{i}-\vec{r}_{j}|^{k}$. 
The simplest case is a system of particles or
clusters thermalized in an harmonic oscillator which is suddenly taken off,
and freely evolving in the vacuum afterwards. In this case the system
experiences a self similar expansion at any successive time, and the exact
time dependent density matrix is given by an isobar canonical Gibbs
equilibrium in the local rest frame. This schematic example may have some
relevance in the study of low energy nuclear collisions, where the so called
freeze out hypothesis\cite{bondorf} describes the strongly interacting
diluted nuclear system as an ideal gas of clusters\cite{fisher}. The presented 
formalism provide a systematique treatment of the confinement and of the successive 
flow.

\begin{acknowledgments}
Discussions with A.Bonasera are gratefully acknowledged.
\end{acknowledgments}

\section{Appendix A: Higher multipoles deformations}

The reasoning of section \ref{deformed} can be extended to higher multipoles
of rank $k$, typically considering the operators $\hat{B}_{i}^{(0)}=\sum_{n}
\hat{r}_{i,n}^{k}$, where $\hat{r}_{i,n}$ is the $i^{th}$ component of the
coordinate of the particle $n$. Then to have a complete information on the
system at any time we have to introduce $k$ additional observables $\hat{B}
_{i}^{(l)},l=1,\dots ,k$ given by 
\[
\hat{B}_{i}^{(l)}=\frac{(-1)^{l}}{\left( 2m\right) ^{l}}\frac{k!}{\left(
k-l\right) !}\sum_{n}\sum_{s=0}^{l}\frac{l!}{s!(l-s)!}
p_{i,n}^{s}r_{i,n}^{k-l}p_{i,n}^{l-s} 
\]
This produces non Hubblian flows and non Maxwellian local momentum
distributions. For example a $\hat{B}^{(0)}=\sum_{n}\hat{x}_{n}^{4}$
tensorial deformation, which is part of an hexadecapole constraint,
introduces the time dependent constraints $\hat{B}_{x}^{(1)}=-2\sum_{n}(\hat{
p}_{x_{n}}\hat{x}_{n}^{3}+\hat{x}_{n}^{3}\hat{p}_{x_{n}})/m$ , $\hat{B}
_{x}^{(2)}=3\sum_{n}(\hat{p}_{x_{n}}^{2}\hat{x}_{n}^{2}+2\hat{p}_{x_{n}}\hat{
x}_{n}^{2}\hat{p}_{x_{n}}+\hat{x}_{n}^{2}\hat{p}_{x_{n}}^{2})/m^{2}$ , $\hat{
B}_{x}^{(3)}=-3\sum_{n}(\hat{p}_{x_{n}}^{3}\hat{x}_{n}+3\hat{p}_{x_{n}}^{2}
\hat{x}_{n}\hat{p}_{x_{n}}+3\hat{p}_{x_{n}}\hat{x}_{n}\hat{p}_{x_{n}}^{2}+
\hat{x}_{n}\hat{p}_{x_{n}}^{3})/m^{3}$ , $\hat{B}_{x}^{(4)}=$ $
24\sum_{n}p_{n}^{4}/m^{4}$. The first term can be interpreted as a cubic
flow meaning that the expansion is non self-similar. The second term can be
seen as a local reorganization of the temperature while the last one is an
explicit deviation from\thinspace the Gaussian distribution of velocities
i.e. from a Maxwell distribution as expected from the Boltzmann factor of an
ideal gas. The generalization to several dimensions, i.e. to constraints of
the type $\hat{B}^{(0)}=\sum_{n}\hat{x}_{n}^{k_{x}}\hat{y}_{n}^{k_{y}}\hat{z}
_{n}^{k_{z}}$ is tedious but straightforward. The $\hat{B}^{(l)}$ are linear
combinations of products of $\hat{p}_{x}^{i}$ , $\hat{p}_{y}^{j}$ and $\hat{p
}_{z}^{k}$ with $\hat{x}^{i^{\prime }},$ $\hat{y}^{j^{\prime }}$ and $\hat{z}
^{k^{\prime }}$, with $i,j,k=0,\dots ,l$ and $k_{x}=i^{\prime }+i,$ $
k_{j}=j^{\prime }+j$ and $k_{z}=k^{\prime }+k$.

It is interesting to notice that considering a generic constraint $\hat{B}
^{(0)}=\sum_{n}f(\vec{\hat{r}}_{n})$, the first correction in time is $
\hat{B}^{(1)}=-\frac{1}{2m}\sum_{n}\vec{\hat{p}}_{n}\cdot \vec{\nabla }
f(\vec{\hat{r}}_{n})+\vec{\nabla }f( \vec{\hat{r}}_{n})$ $\cdot 
\vec{\hat{p}}_{n}$ so that for short time fluctuations $(t-t_{0})=\delta
t $ the statistical ensemble reads (at the first order in $\delta t)$ 
\[
\hat{D}_{\beta ,\lambda }(t)=\frac{1}{Z_{\beta ,\lambda }}\exp
\sum_{n}-\beta \left( \frac{\left( \vec{\hat{p}}_{n}-\vec{A}\left( 
\vec{\hat{r}}_{n}\right) \right) ^{2}}{2m}+U\left( \vec{\hat{r}}
_{n}\right) \right) 
\]
with 
\begin{eqnarray*}
\vec{A}\left( \vec{\hat{r}}_{n}\right) &=&\delta t\vec{\nabla }U( 
\vec{\hat{x}}_{n}) \\
U\left( \vec{\hat{r}}_{n}\right) &=&\frac{\lambda }{\beta } f(\vec{
\hat{x}}_{n})
\end{eqnarray*}
which is akin to an equilibrium of particles in the external scalar and
vector field $\hat{U}$ and $\vec{\hat{A}}$.

\section{Appendix B: Real Gas}

\smallskip Let us now study the case of a real gas with $\hat{H}=\hat{K}+
\hat{V}$ with $\hat{V}=\sum_{nn^{\prime }}V(\hat{r}_{nn^{\prime }})$ a two
body interaction depending only upon the relative distance $\hat{r}
_{nn^{\prime }}=\left| \vec{\hat{r}}_{n}-\vec{\hat{r}}_{n^{\prime
}}\right| .$ Let us investigate the minimum biased distribution including a
compactness observable $\vec{\hat{R}}^{2}$ known at a time $t_{R}.$ $V$
depending only upon $\hat{r}$ , $\hat{B}_{R}^{(1)}$ remains unchanged: $\hat{
B}_{R}^{(1)}=\{\hat{K},\vec{\hat{R}}^{2}\}=-\frac{1}{m}\sum_{n}\vec{
\hat{p}}_{n}\cdot \vec{\hat{r}}_{n}+\vec{\hat{r}}_{n}\cdot \vec{
\hat{p}}_{n}$ while $\hat{B}_{R}^{(2)}$ contains an additional term 
\begin{eqnarray*}
\{\hat{U},\hat{B}_{R}^{(1)}\} &=&
%-\frac{1}{m}\sum_{nn^{\prime }}\left\{ V(%
%\hat{r}_{nn^{\prime }}),\left( \vec{\hat{p}}_{n}\cdot \vec{\hat{r}}%
%_{n}+\vec{\hat{r}}_{n}\cdot \vec{\hat{p}}_{n}\right) \right\}=
-\sum_{nn^{\prime }}\frac{2\vec{\hat{r}}_{n}\cdot \vec{\nabla }V(\hat{%
r}_{nn^{\prime }})}{m} 
%\\
%&=&
=-\sum_{nn^{\prime }}\frac{1}{m}\hat{r}_{nn^{\prime }}\vec{\partial }%
_{r}V(\hat{r}_{nn^{\prime }})
\end{eqnarray*}
and thus reads 
\[
\hat{B}_{R}^{(2)}=\sum_{n}2\vec{\hat{p}}_{n}^{2}/m^{2}-\sum_{nn^{\prime }}%
\frac{1}{m}\hat{r}_{nn^{\prime }}\vec{\partial }_{r}V(\hat{r}_{nn^{\prime
}}) 
\]
In case of an harmonic interaction the $\hat{B}_{R}^{(p)}$ operators only
contain quadratic terms $\sum_{n}\vec{\hat{p}}_{n}^{2}$, $
\sum_{nn^{\prime }}\hat{r}_{nn^{\prime }}^{2}$ and $\sum_{nn^{\prime }}
\vec{\hat{r}}_{nn^{\prime }}\cdot \vec{\hat{p}}_{nn^{\prime }}$, with $
\vec{\hat{p}}_{nn^{\prime }}=\vec{\hat{p}}_{n}-\vec{\hat{p}}
_{n^{\prime }}$. In this case the time evolution can be taken into account
by a suitable time dependent reorganization of the temperature and the
introduction of a time odd constraint, the radial flow. However, for any
other interaction $\hat{B}_{R}^{(2)}$ modifies not only the temperature but
also the two-body interaction. If we define ${V}^{\prime }(\hat{r})={V}(\hat{
r})+\hat{r}\vec{\partial }_{r}V(\hat{r})/4$ and we work out the third
order term: 
\begin{eqnarray*}
\hat{B}_{R}^{(3)} &=&\frac{2}{m^{2}}\sum_{nn^{\prime }}\left\{ V(\hat{r}%
_{nn^{\prime }}),\vec{\hat{p}}_{n}^{2}\right\} -\frac{1}{2m^{2}}%
\sum_{nn^{\prime }}\left\{ \vec{\hat{p}}_{n}^{2},\hat{r}_{nn^{\prime }}%
\vec{\partial }_{r}V(\hat{r}_{nn^{\prime }})\right\} \\
&=&\frac{1}{m^{2}}\sum_{nn^{\prime }}\vec{\hat{p}}_{nn^{\prime }}\cdot 
\vec{\hat{r}}_{nn^{\prime }}\frac{\vec{\partial }_{r}V^{\prime }(\hat{r%
}_{nn^{\prime }})}{\hat{r}_{nn^{\prime }}}+\frac{\vec{\ \partial }%
_{r}V^{\prime }(\hat{r}_{nn^{\prime }})}{\hat{r}_{nn^{\prime }}}\vec{\hat{%
r}}_{nn^{\prime }}\cdot \vec{\hat{p}}_{nn^{\prime }}
\end{eqnarray*}
we can see that the time dependence of the process induces an effective
momentum dependent two-body interaction. An interesting phenomenon occurs at
the level of the next order. Indeed while the kinetic energy term in the
Hamiltonian leads to a first term in $\hat{B}_{R}^{(4)}=\{\hat{H},\hat{B}
_{R}^{(3)}\}$ 
\[
\{\hat{K},\hat{B}_{R}^{(3)}\}=\frac{1}{2m^{3}}\sum_{nn^{\prime }}\left\{ 
\vec{\hat{p}}_{n}^{2},\vec{\hat{p}}_{n}\cdot \vec{\hat{r}}%
_{nn^{\prime }}\frac{\vec{\partial }_{r}V^{\prime }(\hat{r}_{nn^{\prime
}})}{\hat{r}_{nn^{\prime }}}+\frac{\vec{\partial }_{r}V^{\prime }(\hat{r}%
_{nn^{\prime }})}{\hat{r}_{nn^{\prime }}}\vec{\hat{r}}_{nn^{\prime
}}\cdot \vec{\hat{p}}_{n}\right\} 
\]
which is again a akin to a two body interaction, the interaction part
produces a three-body term 
\begin{eqnarray*}
\{\hat{V},\hat{B}_{R}^{(3)}\} &=&\frac{1}{m^{2}}\sum_{nn^{\prime }n"}\left\{
V(\hat{r}_{nn^{"}}),\vec{\hat{p}}_{n}\cdot \vec{\hat{r}}_{nn^{\prime }}%
\frac{\vec{\partial }_{r}V^{\prime }(\hat{r}_{nn^{\prime }})}{\hat{r}%
_{nn^{\prime }}}+\frac{\vec{\partial }_{r}V^{\prime }(\hat{r}_{nn^{\prime
}})}{\hat{r}_{nn^{\prime }}}\vec{\hat{r}}_{nn^{\prime }}\cdot \vec{%
\hat{p}}_{n}\right\} \\
&=&\frac{2}{m^{2}}\sum_{nn^{\prime }n"}\vec{\partial }_{r}V(\hat{r}%
_{nn^{"}})\vec{\partial }_{r}V^{\prime }(\hat{r}_{nn^{\prime }}) \\
&=&\frac{1}{m^{2}}\sum_{nn^{\prime }n"}\vec{\partial }_{r}V(\hat{r}%
_{nn^{"}})\left( \frac{5}{4}\vec{\partial }_{r}V(\hat{r}_{nn^{\prime }})+%
\frac{\hat{r}_{nn^{\prime }}}{4}\vec{\partial }_{r}^{2}V(\hat{r}%
_{nn^{\prime }})\right)
\end{eqnarray*}

\section{Appendix C: details about the RPA}

In eq.(\ref{rpa:2}) we have expressed the RPA matrix as
$\mathcal{H} = \mathcal{W}-\mathcal{N\,V}$
where the self consistent mean field is defined as 
$\mathcal{W}\parallel \cdot\gg =\parallel \left[ W,\cdot \right] \gg ,$ 
the residual interaction $\mathcal{V}\parallel
\cdot \gg =\parallel \frac{\partial W}{\partial \rho ^{T}}\cdot \gg$
and the density operator $\mathcal{N}\parallel \cdot \gg
=$ $\parallel \left[ \rho ,\cdot \right] \gg $.

To show the hermiticity of these operators 
one should go back to the definition $\ll A\parallel 
\mathcal{B}^{+}\parallel C\gg =\left( \ll C\parallel \mathcal{B}\parallel
A\gg \right) ^{*}.$ In the case of $\mathcal{W}$ and $\mathcal{N}$ ,
using $\left( TrC^{+}\left[ B,A\right] \right) ^{*}=TrA\left[ B,C\right] $
valid when $B$ is hermitian $B^{+}=B$, we get $\ll A\parallel \mathcal{B}
^{+}\parallel C\gg =\ll A\parallel \mathcal{B}\parallel C\gg .$ Concerning
the residual interaction, using the hermiticity of $\hat{W}$ and its
definition we can easily get an explicitly hermitian form $\mathcal{V}
_{[1,2]}=\partial ^{2}E/\partial \hat{\rho}_{\left[ 1\right] }^{*}\partial 
\hat{\rho}_{\left[ 2\right] }^{T}.$ 

Then we have expressed the generalized constraints of the 
time dependent RPA problem in the small amplitude limit as 

\begin{eqnarray}
\hat{A}_{\ell }&=&a_{\upsilon _{\ell }}\hat{C}_{\upsilon _{\ell }}+a_{\upsilon
_{\ell }}^{*}\hat{C}_{\upsilon _{\ell }}^{+} \\
 \hat{B}_{\ell }&=&i\omega _{\nu _{\ell }}a_{\upsilon _{\ell }}\hat{C}%
_{\upsilon _{\ell }}-i\omega _{\nu _{\ell }}a_{\upsilon _{\ell }}^{*}\hat{C}%
_{\upsilon _{\ell }}^{+} \label{collective}
\end{eqnarray}

and we have interpreted $\hat{A}_{\ell }$ 
as a collective coordinate associated
with the creation of a collective mode through the $\hat{C}_{\upsilon _{\ell
}}^{+}$ operator, and $\hat{B}_{\ell }$ as the associated momentum. 

Let us justify this interpretation.

%Three properties of the RPA matrix are important.
If $\omega _{\nu }$ is an eigenvalue of the RPA matrix correponding to 
the eigenstate $\delta \hat{\rho}_{\nu }$, 
it is easy to show that $\delta \hat{\rho}_{\nu }^{+}$ is solution of 
$\omega _{\nu }^{*}\parallel \delta \rho _{\nu }^{+}\gg =-\mathcal{H}
\parallel \delta \rho _{\nu }^{+}\gg $ so that $\parallel \delta \rho _{\nu
}^{+}\gg $ is also an eigenstate of the RPA matrix associated to $-\omega
_{\nu }^{*}.$ Thus RPA solutions can be grouped by pairs.

Then, because of the commutator structure of the dynamical equation, the RPA
equation do not propagates the diagonal terms $\delta \hat{\rho}_{II}$, so
we can focus on the off diagonal terms 
%$I\neq J$ and $K\neq L$
. Thus we
can introduce a collective operator $Q$ such that $\parallel \delta \rho \gg
=\parallel [\rho ,Q]\gg =\mathcal{N}\parallel Q\gg .$ In fact the density
variation $\delta \rho $ can be interpreted as produced by a unitary
transformation generated by the operators $Q$ and $Q^{+}$: $\hat{\rho}
\rightarrow e^{-i\lambda (\hat{Q}+\hat{Q}^{+})}\hat{\rho}e^{i\lambda (\hat{Q}
+\hat{Q}^{+})}=\hat{\rho}+i\lambda ([\rho ,Q]+[\rho ,Q^{+}]).$ Thus the
relation holds $\omega _{\nu }\parallel Q_{\nu }\gg =\mathcal{N}^{-1}
\mathcal{HN}\parallel Q_{\nu }\gg .$

If $\left[ W,\rho \right] =0,$ then $\mathcal{N}^{-1}\mathcal{HN=\mathcal{
W}-\mathcal{\,V\mathcal{N}=}H}^{+}$ meaning that $\parallel Q_{\nu }\gg $ is
an eigen vector of $\mathcal{H}^{+}$ associated with the eigenvalue $\omega
_{\nu }.$This means that, when $[W,\rho ]=0,$ $\omega _{\nu }^{*}$ is also
an eigenvalue of the RPA\ matrix and, since the RPA eigenvalues appears by
pairs $-\omega _{\nu }$ is also a solution. Requiring the equality of these
to solutions with the pair $\omega _{\nu }$ and -$\omega _{\nu }^{*}$ shows
that $\omega _{\nu }$ is real or purely imaginary and is associated with a
second solution $-\omega _{\nu }.$

If we now introduce the dual basis $\ll C_{\upsilon
}\parallel $ such that $\ll C_{\upsilon ^{\prime }}\parallel \delta \rho
_{\nu }\gg =\delta _{\nu ^{\prime }\nu }$ and the associated closure
relation $\sum_{\nu }\parallel \delta \rho _{\nu }\gg \ll C_{\upsilon
}\parallel =1$, the relation $\ll C_{\upsilon }\parallel \mathcal{H=}\omega
_{\nu }\ll C_{\upsilon }\parallel $ is easily demonstrated. 
Then $\omega _{\nu }^{*}\parallel C_{\nu }\gg =$ $\mathcal{H
}^{+}\parallel C_{\nu }\gg $ also holds showing that, when $\left[ W,\rho
\right] =0,$ $\hat{C}_{\nu }=\hat{Q}_{\upsilon }$ if $\omega _{\nu }$ is
real.

This shows that the dual of the eigenvector $\parallel \delta \rho
_{\nu }\gg$ is the operator responsible of the excitation of the collective
mode $\omega _{\nu }$ as indicated in equation (\ref{collective})
with $a_{\upsilon _{\ell }}=\ll A_{\ell }\parallel
\delta \hat{\rho}_{\nu }\gg$.

\end{document}